\documentclass[12pt]{JHEP3}
\usepackage{mathrsfs}
\usepackage{amsmath,amssymb}
\usepackage{epsfig}
\usepackage{relsize}
\usepackage{graphicx}

\def\x{{\tilde x}}

\def\sl(2){\alg{sl}(2)}

\def\be{\begin{equation}}
\def\ee{\end{equation}}

\newcommand{\bea}{\begin{eqnarray}}
\newcommand{\eea}{\end{eqnarray}}

\def\s {\sigma}

\def\om {\omega}

\def\la{\label}

\def\ov{\over}

\def\vp{\varphi}

\newcommand{\alg}[1]{\mathfrak{#1}}
\newcommand{\su}{\alg{su}}

\newcommand{\psu}{\alg{psu}}

\newcommand{\AdS}{{\rm  AdS}_5\times {\rm S}^5}

\newcommand{\atopfrac}[2]{\genfrac{}{}{0pt}{}{#1}{#2}}

\newcommand{\bem}{\left (\begin{matrix}}
\newcommand{\eem}{\end{matrix} \right )}

\def\bei{\begin{itemize}}
\def\eei{\end{itemize}}

\author{Gleb Arutyunov$^a$\footnote{Email: G.E.Arutyunov@uu.nl, frolovs@maths.tcd.ie} {}\footnote{Correspondent fellow at Steklov
Mathematical Institute, Moscow.}\, and\,  Sergey Frolov$^{b\, \dagger}$
 \\ $^{a}$ {\it Institute for Theoretical
Physics and Spinoza Institute,\\ ~~Utrecht University, 3508 TD
Utrecht, The Netherlands} \\ $^b$ {\it Hamilton Mathematics Institute and School of Mathematics, \\
~~Trinity College, Dublin 2,
Ireland} }

\abstract{ We utilize the DHM integral representation for the BES
dressing factor of  the world-sheet S-matrix of the $\AdS$
light-cone string theory, and the crossing equations to fix the
principal branch of the dressing factor on the rapidity torus. The
results obtained are further used,  in conjunction with the fusion
procedure, to determine the bound state dressing factor of the
mirror theory. We convincingly demonstrate that the mirror bound
state S-matrix found in this way does not depend on the internal
structure of a bound state solution employed in the fusion
procedure. This welcome feature is in perfect parallel to string
theory, where the corresponding bound state S-matrix has no
bearing on bound state constituent particles as well. The mirror
bound state S-matrix we found provides the final missing piece in
setting up the TBA equations for the $\AdS$ mirror theory.

}

\title{The Dressing Factor and Crossing Equations
}

\preprint{%\smaller{\smaller{\smaller{???}}}\\[-.5ex]
          \smaller{\smaller{\smaller{ITP-UU-09-17}}}\\[-.5ex]
          \smaller{\smaller{\smaller{SPIN-09-17}}}\\[-.5ex]
          \smaller{\smaller{\smaller{TCDMATH-09-12}}}\\[-.5ex]
 \smaller{\smaller{\smaller{HMI-09-06}}}}
\begin{document}

\renewcommand{\thefootnote}{\arabic{footnote}}
\setcounter{footnote}{0}
%%%%%%%%%%%%%%%%%%%%%%%%%%%%%%%
%%%%%%%%%%%%%%%%%%%%%%%%%%%%%%%%%

%\vskip 0.5cm

\section{Introduction}
The exact finite-size spectrum of the $\AdS$ superstring remains
one of the most important challenges in the AdS/CFT correspondence
\cite{M}. In the light-cone gauge the string sigma model is
formulated on a two-dimensional cylinder of finite circumference
proportional to the light-cone momentum. When the light-cone
momentum tends to infinity, the string world-sheet decompactifies
and, under the assumption of quantum integrability of the
light-cone model, one can apply factorized scattering theory
\cite{ZZ} and the Bethe ansatz to capture the spectrum. A lot of
remarkable progress has been achieved in this way, for the recent
reviews and extensive list of references, see, e.g.
\cite{AFrev}-\cite{Rej:2009je}.

\smallskip

On the infinite world-sheet the string sigma model exhibits
massive excitations which transform in the tensor product of two
fundamental representations of the $\psu(2|2)$ superalgebra
enhanced by two central charges, both dependent on the generator
of the world-sheet momentum \cite{B,AFPZ}. The symmetry algebra
severely constrains the matrix form of the two-particle S-matrix
\cite{B} and, together with the Yang-Baxter equation and the
requirement of generalized physical unitarity \cite{AFZ,AFtba},
leads to its unique determination up to an overall scalar function
of particle momenta. This function includes, as its most
non-trivial piece, the so-called dressing factor
$\sigma=\exp(i\theta)$, where $\theta$ is the dressing phase
\cite{AFS}. Taking into account that the light-cone S-matrix is
compatible with the assumption of crossing symmetry, one finds
that the latter implies certain functional relations for the
dressing factor, known as the crossing equations \cite{Janik}. A
solution to these equations has been found in the strong coupling
expansion in terms of an asymptotic series \cite{BHL}. The
corresponding dressing factor appears to be compatible with all
available classical and perturbative string data  \cite{AFS},
\cite{KMMZ}-\cite{Casteill:2007ct} and it passed a number of
further non-trivial checks, see {\it e.g.} \cite{G1,G2}.

\smallskip

The dressing factor suitable for the weak coupling expansion has
been also proposed by assuming a certain analytic continuation of
its strong-coupling counterpart \cite{BES}, and in what follows we
refer to it as the Beisert-Eden-Staudacher (BES) dressing factor.
The proposal was successfully confronted against direct
field-theoretic perturbative calculations \cite{Bern}-\cite{BMR}.
However, it has not been shown so far that the BES dressing factor
satisfies crossing equations for finite values of the string
coupling. Filling this gap is among the goals of this paper.

\smallskip

Another important issue which motivates our present effort is
related to the finite-size spectral problem. When the light-cone
momentum is finite, the description of the string spectrum in
terms of the Bethe equations \cite{S,BS} fails, and one has to
resort to new methods to find a solution of  the corresponding
spectral problem. In this respect an exciting possibility is
offered by the Thermodynamic Bethe Ansatz approach (TBA)
originally developed for relativistic integrable models \cite{za}.
In essence, the TBA allows one to relate the finite-size ground
state energy in a two-dimensional integrable model with the free
energy (or, depending on the boundary conditions, with Witten's
index) in the so-called mirror model that is obtained from the
original model by a double Wick rotation. A peculiarity of the
non-relativistic case, as provided by the string sigma model, is
that the original and the mirror Hamiltonians are not the same.
This brings certain subtleties in setting up the corresponding TBA
approach \cite{AFtba}.

\smallskip

Recently, we have derived a set of the TBA equations for the
$\AdS$ mirror model \cite{AFsh,AFmtba}. A parallel development
took place in the works \cite{GK1,GK2} and
\cite{Bombardelli:2009ns}, where also the so-called Y-system has
been proposed. The Y-system represents a set of local equations
which is believed to contain the whole spectrum (the ground and
excited states), and it is obtained from the original infinite set
of TBA equations by taking a certain projection. In one respect
the undertaken derivations of the mirror TBA equations and the
associated Y-system remain incomplete. To completely define the
TBA equations and, therefore, to make them a working device, one
has to understand the so far unknown analytic properties of the
dressing factor in the kinematic region of the mirror theory.

\smallskip

To be precise, the BES dressing phase is represented by a double
series convergent in the region $|x_{1}^{\pm}|>1$ and
$|x_2^{\pm}|>1$, where $x_{1,2}^{\pm}$ are kinematic parameters
related to the first and the second particle, respectively. This
series admits an integral representation found by Dorey, Hofman
and  Maldacena (DHM) \cite{DHM} which is valid in the same region
of kinematic parameters and for finite values of the string
coupling.\footnote{Another useful integral representation was
found in \cite{SKV}. } To find the BES dressing phase in other
kinematic regions of interest (in the mirror region), one has to
analytically continue the DHM integral representation beyond
$|x_{1,2}^{\pm}|>1$. Understanding this continuation is precisely
the subject of this paper.

\smallskip

Regarding the dressing phase as a multi-valued function, we want
to fix its particular analytic branch which satisfies the crossing
equations. In fact, continuation compatible with crossing is the
main rationale behind our treatment. Consideration of crossing
requires us to associate to each of the particles (bound states)
the so-called rapidity torus (the $z$-torus) which uniformizes the
corresponding dispersion relation \cite{Janik}. The analytically
continued dressing phase should be then understood as a function
$\theta(z_1,z_2)$, where $z_1$ and $z_2$ belong to the
corresponding tori, see section 2 for a more precise definition.
This function must satisfy the crossing equation in each of its
arguments for $z_1,z_2$ being anywhere on the  tori. Thus, to
construct $\theta(z_1,z_2)$, we choose an analytic continuation
path on the $z$-torus which starts in the particle region, where
$|x^{\pm}(z)|>1$, and penetrates the other regions of the
$z$-torus. Following this path, we properly account for the change
of the DHM integral representation to guarantee the continuity of
our resulting function. In this way we build up the analytic
continuation of the dressing phase to all possible kinematic
regions and then verify the fulfillment of the crossing equations.

\smallskip

We further study the analytic continuation of the dressing factor
for bound states of the string model. This dressing factor is
obtained by fusing the dressing factors of bound state constituent
particles, all of them being in the region $|x^{\pm}(z)|>1$ of the
elementary $z$-torus. The dressing factor is then analytically
continued into the whole bound state $z$-torus in a way compatible
with the corresponding crossing equations.

\smallskip

In view of applications to the TBA program, we also determine the
dressing factor for bound states of the mirror model by fusing the
dressing factors of constituent mirror particles. Since some of
these particles fall necessarily outside the region
$|x^{\pm}(z)|>1$, in the fusion procedure we use their
analytically continued expressions. Equations defining a
$Q$-particle bound state admit $2^{Q-1}$ different solutions and,
since any of these solutions can be used in the fusion procedure,
this raises a question about uniqueness of the final result. To
find an answer, it is important to realize that dealing with
mirror TBA equations, we are mainly interested not in the bound
state dressing factor $\sigma^{QQ'}$ itself but rather in the
following quantity \bea\nonumber \Sigma^{QQ'} =
\sigma^{QQ'}\,\prod_{j=1}^Q\prod_{k=1}^{Q'} {1-{1\ov x_j^+
z_k^-}\ov 1-{1\ov x^-_j z^+_k}}\eea which logarithmic derivative
appears as one of the TBA kernels \cite{AFmtba}. Here $x_j^{\pm}$
and $z_k^{\pm}$ are the kinematical parameters of the constituent
particles corresponding to $Q$- and $Q'$-particle bound states,
respectively. The {\it improved} factor $\Sigma^{QQ'}$ originates
from fusion of the scalar factors of mirror theory scattering
matrices corresponding to bound state constituent particles. By
evaluating $\Sigma^{QQ'}$ on a particular bound state solution, we
then show that the resulting expression depends on the bound state
kinematic parameters only and that all the dependence on
constituent particles completely disappears! This indicates that
$\Sigma^{QQ'}$ is the same for all bound state solutions, as we
also confirm by explicitly evaluating it on yet another solution.
In this respect $\Sigma^{QQ'}$ appears as good as the
corresponding bound state dressing factor of the original string
theory. We derive  an explicit formula (\ref{sigtot3}) for
$\Sigma^{QQ'}$ which can be used to complete the mirror TBA
equations.

\smallskip

The paper is organized as follows. In the next section we recall
the necessary facts about the dressing phase, rapidity torus and
the crossing equations for both fundamental particles and the
bound states. In section 3 we introduce our basic building blocks
-- the $\Phi$ and $\Psi$-functions -- and study their analytic
properties. In section 4 we determine a particular analytic
continuation of the dressing phase for fundamental particles and
verify that it obeys the crossing equations. In section 5 we do
the same for bound states of string theory. In principle, the
material of section 5 contains the one of section 4, as
fundamental particles can be regarded as one-particle bound
states. However, the discussion in section 4 gives a clear picture
of how the analytic continuation is constructed and, therefore,
provides a good preparation for understanding the technically more
involved issue of the bound states. In section 6 we determine the
bound state dressing factor of the mirror theory and argue that it
has the same universal form regardless of a bound state solution
chosen for its construction. Finally, we conclude by discussing
the results obtained. In four appendices we collected some
identities satisfied by $\Psi$-functions, as well as technical
details of the derivations presented in the main body of the
paper.

\newpage
%%%%%%%%%%%%%%%%%%%%%%%%%%%%%%%%%%%%
\section{The dressing phase and crossing equations}\la{dresscross}

In the light-cone gauge the string sigma model exhibits a massive
spectrum. The corresponding particles transform in the tensor
product of two fundamental multiplets of the centrally extended
$\su(2|2)$ superalgebra. Besides the fundamental particles the
asymptotic spectrum also contains their bound states which
manifest themselves as poles of the scattering matrix for
fundamental particles. A $Q$-particle bound state
\cite{Dorey:2006dq} transforms in the tensor product of two
$4Q$-dimensional atypical totally symmetric multiplets of the
centrally extended $\su(2|2)$ superalgebra
\cite{Beisert:2006qh,Chen:2006gp}. Both the fundamental and the
bound state S-matrices are determined by kinematical symmetries
and additional physicality requirements up to an overall scalar
factor. Upon normalizing the kinematically determined S-matrix in
a certain (canonical) way, the corresponding scalar factor
acquires an absolute meaning and therefore becomes an important
dynamical characteristic of the model. For the $\AdS$ superstring
the scalar factor is proportional to the dressing factor $\sigma$,
the latter can be expanded over local conserved charges of the
model \cite{AFS}. By unitarity $\sigma=\exp(i\theta)$, where
$\theta$ is known as the {\it dressing phase}. This section has an
introductory character, its purpose is to recall the relevant
facts about the dressing phase as well as to introduce the
necessary notation.

\subsection{Uniformization tori}
As was argued in \cite{Janik}, the world-sheet S-matrix appears to
be compatible with an assumption of crossing symmetry that
corresponds to replacing a particle for its anti-particle.
Implementation of this symmetry leads to a non-trivial functional
relation for the dressing phase.

\smallskip

Similar to the more familiar relativistic case, a rigorous
treatment of crossing symmetry requires finding a Riemann surface
which uniformizes the dispersion relation of the model. For
fundamental particles this uniformization is achieved by
introducing a torus with real and imaginary periods given by
\bea\la{periods} 2\omega_1=4{\rm K}(k)\, , ~~~~~~~~~
2\omega_2=4i{\rm K}(1-k)-4{\rm K}(k)\, ,
 \eea
respectively \cite{Janik,Beisert:2006zy} . Here ${\rm K}(k)$
stands for the complete elliptic integral of the first kind. The
elliptic modulus $k=-4g^2$, where $g$ is the string tension
related to the `t Hooft coupling $\lambda$ as
$g=\frac{\sqrt{\lambda}}{2\pi}$.

%\vskip 0.7cm \noindent
%---------- FIGURE TOP ------------
\begin{figure}
\begin{minipage}{\textwidth}
\begin{center}
\includegraphics*[width=0.79\textwidth]{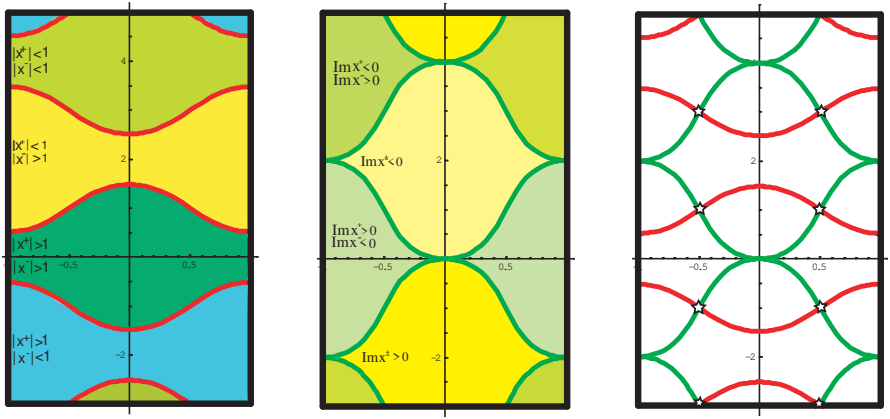}
\parbox{5in}{\caption{ \label{torus1} On the left figure the torus is divided
by the curves $|x^+|=1$ and $|x^-|=1$ into four non-intersecting
regions. The middle figure represents the torus divided by the
curves ${\rm Im}(x^+)=0$ and ${\rm Im}(x^-)=0$, also in four
regions. The right figure contains all the curves of interest.
 }}
\end{center}
\end{minipage}
\end{figure}
%---------- FIGURE END ------------

\smallskip

Analogously, the dispersion relation corresponding to a
$Q$-particle bound state is also uniformized by an elliptic curve
whose periods are obtained by replacing in eq.(\ref{periods}) the
coupling constant $g$  with $g/Q$. Neglecting for the moment the
dressing factor, the {\it canonically} normalized S-matrix  that
describes scattering of $Q$- and $M$-particle bound states turns
out to be a meromorphic matrix function $S_{QM}(z_1,z_2)$ defined
on the product of the corresponding tori \cite{AFtba,AFbs}. In
what follows we will refer to a torus supporting fundamental or
bound state representations as the $z$-torus.

\smallskip

In addition, we will also need the variables $x^{\pm}$ which
satisfy the following constraint \bea\la{const1} x^++{1\ov
x^+}-x^--{1\ov x^-}= {2i\ov g}\,  \eea and which are related to
the torus variable $z$ through $\frac{x^+}{x^-}=\exp(i\, 2 {\rm
am}\, z)$, where ${\rm am}\, z$ is the Jacobi amplitude. The
functions $x^{\pm}(z)$ are meromorphic on the $z$-torus
corresponding to fundamental particles. The torus itself can be
divided into four non-intersecting regions in two different ways:
either by curves $|x^{\pm}|=1$ or by curves ${\rm Im}\,
x^{\pm}=0$, see Figure 1. These divisions will play an
important role in our subsequent discussion of the dressing phase.

\smallskip

Under the crossing symmetry transformation the variables $x^{\pm}$
undergo the inversion: $x^{\pm}\to 1/x^{\pm}$.  In terms of
variable $z$, this map is realized as a shift by the imaginary
half-period: $z\to z\pm \omega_2$ \cite{Janik}. The dressing
phase, however, is not periodic on the $z$-torus and should be
considered as a function on the product of two $z$-planes. The
crossing symmetry allows one to define the dressing phase on the
product of two complex planes from its knowledge on the $z$-torus,
the latter being regarded as the fundamental domain. As we will
see later on, inside this fundamental domain the dressing phase
has an infinite number of cuts.

\subsection{Dressing and crossing for fundamental particles}

The dressing phase for fundamental particles is a function of the
constrained variables $x^{\pm}$: $\theta\equiv
\theta(x_1^+,x_1^-,x_2^+,x_2^-)$ and, in the region
$|x_{1,2}^{\pm}|>1$, it admits the following double series
expansion \cite{AFS,BK}
 \bea
 \la{AFS}
\theta(x_1^+,x_1^-,x_2^+,x_2^-) = \sum_{r=2}^\infty
\sum_{\textstyle\atopfrac{s>r}{ r+s={\rm odd}} }^\infty
c_{r,s}(g)\Big[ q_r(x_1^\pm)q_{s}(x_2^\pm) -
q_r(x_2^\pm)q_{s}(x_1^\pm) \Big]\, , \la{phase}
 \eea
where the local conserved charges $q_r(x^\pm)$ are \bea
\la{localconcharge} q_r(x_k^-,x_k^+) &=& {i\ov r - 1}\left[
\left({1\ov x^+_k}\right)^{r - 1} - \left({1\ov x^-_k}\right)^{r -
1}\right]\, . \eea Here the coefficients $c_{r,s}(g)$ are
non-trivial real functions of the string tension  admitting  a
well-defined asymptotic expansion for large $g$. The double series
representation (\ref{phase}) implies that the dressing phase in
this region can be written \cite{Arutyunov:2006iu} via a single
skew-symmetric function $\chi$ of two variables  \bea\la{chi}
\chi(x_1,x_2)=-\sum_{r=2}^\infty \sum_{\textstyle\atopfrac{s>r}{
r+s={\rm odd}} }^\infty
\frac{c_{r,s}(g)}{(r-1)(s-1)}\Big[\frac{1}{x_1^{r-1}x_2^{s-1}}-\frac{1}{x_2^{r-1}x_1^{s-1}}\Big]\,
\eea as a sum of four terms  \bea\la{BES}
\theta(x_1^+,x_1^-,x_2^+,x_2^-) =\chi(x_1^+,x_2^+)
-\chi(x_1^+,x_2^-)-\chi(x_1^-,x_2^+)+\chi(x_1^-,x_2^-)\, .
~~~~~\eea  At any given order in the asymptotic $1/g$ expansion
the double series defining $\chi$ is convergent for $|x_{1,2}|>1$.
For the $z$-torus in Figure 1, conditions $|x^{\pm}(z)|>1$ single
out a green region with the shape of a ``fish". Thus, every term
in the strong coupling asymptotic expansion of the dressing phase
$\theta(z_1,z_2)$ is a well-defined function provided both $z_1$
and $z_2$ belong to the ``fish". The set of points on the
$z$-torus obeying $|x^{\pm}(z)|>1$ will be called the particle
region.

\smallskip

Assuming the functional form (\ref{AFS}), the equations implied by
crossing symmetry (to be discussed below) can be solved
perturbatively in the strong coupling expansion, the coefficients
$c_{r,s}(g)$ emerge in the form of an {\it asymptotic series}
\cite{BHL}. On the other hand, the coefficients $c_{r,s}(g)$ admit
a {\it convergent} small $g$ expansion
$$
c_{r,s}(g)=g\sum_{n=r+s-3}^\infty g^n c_{r,s}^{(n)}\, .
$$
The weak coupling coefficients $c_{r,s}^{(n)}$ have been
determined from their strong coupling cousins by assuming a
certain analytic continuation procedure \cite{BES}.

\smallskip

By using the weak coupling expressions for $c_{r,s}(g)$ and the
series (\ref{chi}), in the work \cite{DHM} an integral
representation for $\chi$ valid for finite values of $g$ was
obtained. It is given by the following double integral \bea\nonumber
\chi(x_1,x_2)=i\oint\frac{{\rm d}w_1}{2\pi i}\oint \frac{{\rm
d}w_2}{2\pi i}\frac{1}{(w_1-x_1)(w_2-x_2)} \log{\Gamma\big[1+{i\ov
2}g\big(w_1+\frac{1}{w_1}-w_2-{1\ov w_2}\big)\big]\ov
\Gamma\big[1-{i\ov 2}g\big(w_1+\frac{1}{w_1}-w_2-{1\ov
w_2}\big)\big]}\, , \\ \la{chip} \eea where integrations are
performed over the unit circles. The integral representation
(\ref{chip}) holds for $|x_{1,2}|>1$ and, therefore, from the
point of view of the $z$-torus,  for any finite value of $g$, it
renders the dressing phase (\ref{BES}) a well-defined function on
the product of two particle regions. We will call eq.(\ref{chip})
the DHM integral representation.

\smallskip
Assuming that the dressing phase $\theta(z_1,z_2)$ is defined on
the whole $z$-torus, the functional equations implied by crossing
symmetry read as \bea\la{cr1a}
&&\theta(z_1,z_2)+\theta(z_1+\om_2,z_2) ={1\ov i} \log \Big[{x_2^-\ov x_2^+} h(x_1,x_2)\Big]\,,~~~~\\
\la{cr1b} &&\theta(z_1,z_2)+\theta(z_1,z_2-\om_2) ={1\ov i} \log
\Big[{x_1^+\ov x_1^-}h(x_1,x_2)\Big]\,,~~~~ \eea where \bea\la{h}
h(x_1,x_2)=\frac{x_1^-
-x_2^+}{x_1^--x_2^-}\frac{1-\frac{1}{x_1^+x_2^+}}{1-\frac{1}{x_1^+x_2^-}}\,
.~~~~ \eea Under the double crossing one gets, for instance, \bea
\theta(z_1+2\om_2,z_2)-\theta(z_1,z_2)=\frac{1}{i}\log h_{\cal
D}(x_1,x_2)\, , \la{dcross}\eea where \bea\la{hD} h_{\cal
D}(x_1,x_2)=\frac{(x_1^+-x_2^+)(x_1^--x_2^-)}{(x_1^+-x_2^-)(x_1^--x_2^+)}\frac{\Big(1-\frac{1}{x_1^-x_2^+}\Big)
\Big(1-\frac{1}{x_1^+x_2^-}\Big)}{\Big(1-\frac{1}{x_1^+x_2^+}\Big)
\Big(1-\frac{1}{x_1^-x_2^-}\Big)}\, , \eea which shows that
$2\omega_2$ is not a period of the dressing phase.

\smallskip

It should be stressed that so far crossing symmetry  has been
imposed on the dressing phase in the asymptotic sense only, and
this led, through the analytic continuation procedure from strong
to weak coupling, to the finite $g$ representation defined with
the help of (\ref{chip}). Verification of crossing symmetry for
finite $g$, {\it i.e.} not in the asymptotic sense, constitutes an
open problem. Its solution relies on finding an analytic
continuation of the dressing phase from the particle region to the
whole $z$-torus which is compatible with crossing symmetry.

\smallskip

Our final remark  concerns the integral representation
(\ref{chip}). Formula (\ref{chip}) holds for $|x_{1,2}|>1$ but, in
fact, it can be used to continue the function $\chi$ for each of
its arguments slightly inside the unit circle. Indeed, the
integration contours can be chosen to be circles of any radius
$r>r_{\rm cr}\equiv \sqrt{1+{1\ov 4 g^2}}-{1\ov 2g}$, and,
therefore, the integral representation above can be extended for
$|x_{1,2}|>r_{\rm cr}$. Moreover, if one of the circles in
(\ref{chip}) is of unit radius then the radius of the second
circle can be reduced up to $ \sqrt{1+{1\ov g^2}}-{1\ov g}$.

\subsection{Dressing and crossing for bound states}
The dressing phase which describes scattering of $Q$- and
$M$-particle bound states can be obtained from the dressing phase
for fundamental particles by means of the fusion procedure
\cite{Chen:2006gq,Roiban:2006gs}. The fused dressing phase
$\theta^{QM}$ is given by the same formulae (\ref{BES}) and
(\ref{chip}), where now the variables $x_1^{\pm}$ and $x_2^{\pm}$
are associated to the $Q$- and $M$-particle bound states,
respectively \bea\la{xpxmq} x_1^++\frac{1}{x_1^+}-x^-_1
-\frac{1}{x^-_1}=\frac{2i}{g}Q \, , ~~~~~
x_2^++\frac{1}{x_2^+}-x^-_2 -\frac{1}{x^-_2}=\frac{2i}{g}M\, . \,
\eea

The crossing equations for the dressing factor describing
scattering of $Q$- and $M$-particle bound states are \cite{AFbs}\footnote{The second formula in (\ref{crMq}) follows from the first one by using the unitarity condition $\sigma^{QM}(z_1,z_2)\sigma^{MQ}(z_2,z_1)=1$, and the identity $\prod_{k=0}^{M-1}G(Q-M+2k)=\prod_{k=1}^{Q-1}G(M-Q+2k)$.}
 \bea
 \la{crMq} \begin{aligned}
\sigma^{QM}(z_1,z_2)\sigma^{QM}(z_1+\omega_2,z_2)=\left(\frac{x_2^-}{x_2^+}\right)^Q
\frac{x_1^-
-x_2^+}{x_1^--x_2^-}\frac{1-\frac{1}{x_1^+x_2^+}}{1-\frac{1}{x_1^+x_2^-}}
\prod_{k=1}^{Q-1}G(M-Q+2k)\, ,~~~~
\\
\sigma^{QM}(z_1,z_2)\sigma^{QM}(z_1,z_2-\omega_2)=\left(\frac{x_1^+}{x_1^-}\right)^M
\frac{x_1^--x_2^+}{x_1^--x_2^-}\frac{1-\frac{1}{x_1^+x_2^+}}{1-\frac{1}{x_1^+x_2^-}}
\prod_{k=1}^{Q-1}G(M-Q+2k)\, ,~~~~
\end{aligned}
 \eea where the following function was introduced
$$
G(\ell)=\frac{u_1-u_2-\frac{i}{g}\ell}{u_1-u_2+\frac{i}{g}\ell}\,
.
$$
Here for a $Q$-particle bound state the variable $u\equiv u^Q$ is
defined as
$$
u^Q=x^++\frac{1}{x^+}-\frac{i}{g}Q=x^-+\frac{1}{x^-}+\frac{i}{g}Q\,
.
$$
Taking the logarithm of both sides of eqs. (\ref{crMq}), one obtains
the corresponding equations for the dressing phase. As for the
case of fundamental particles, the dressing phase describing
scattering of bound states is a well-defined function in the
particle region of the $z$-torus and it should be analytically
continued to the whole torus in a way compatible with crossing
equations (\ref{crMq}).

\section{$\Phi$- and $\Psi$-functions}\la{phipsi}
In this section we introduce  $\Phi$- and $\Psi$-functions and
discuss the analytic continuation of $\chi$ in terms of these
functions  for $x_1, x_2$ close to the unit circle.

\subsection{$\Phi$-function}

Let us introduce a function $\Phi(x_1,x_2)$ defined as the
following double integral of the Cauchy type  \bea\la{Phip}
\Phi(x_1,x_2)=i\oint\frac{{\rm d}w_1}{2\pi i}\oint \frac{{\rm
d}w_2}{2\pi i}\frac{1}{(w_1-x_1)(w_2-x_2)} \log{\Gamma\big[1+{i\ov
2}g\big(w_1+\frac{1}{w_1}-w_2-{1\ov w_2}\big)\big]\ov
\Gamma\big[1-{i\ov 2}g\big(w_1+\frac{1}{w_1}-w_2-{1\ov
w_2}\big)\big]}\, , ~~~~ \eea where the integrals are over the
unit circles. Since the integrand is a continuous function of
$\vartheta_1\,,\vartheta_2$ where $w_i = e^{i\vartheta_i}$, the
function $\Phi$ is unambiguously defined by this integral formula
for all values of $x_1,x_2$ not lying on the unit circle.

If $x_1$ (or $x_2$ or both) is on the unit circle, we can define
$\Phi(e^{i\varphi_1},x_2)$ as the following limit  \bea \la{Phi1}
\Phi(e^{i\varphi_1},x_2)\equiv \lim_{\epsilon\to
0^+}\Phi(e^{\epsilon}e^{i\varphi_1},x_2)\,, \eea
 {\it i.e.} we approach the value $x_1=e^{i\varphi_1}$ from exterior of the circle.
 This limit exists because for $|x_1|>1$, $|x_2|\neq 1$, and $|w_2|=1$ the integrand in (\ref{Phip})
 is a holomorphic function of $w_1$ in the annulus $\sqrt{1+{1\ov g^2}}-{1\ov g} < |w_1|<|x_1|$.

If we would define the limiting value of $\Phi$ by approaching
$x_1=e^{i\varphi_1}$ from the interior of the circle we would
obtain the result different from (\ref{Phi1}) and, for this
reason, the function $\Phi$ is not continuous across $|x_1|=1$. In
fact, it is not difficult to see that \bea \la{Phi2}
\lim_{\epsilon\to 0^+}\Phi(e^{-\epsilon}e^{i\varphi_1},x_2)&=&
\Phi(e^{i\varphi_1},x_2) + \\
&& \hspace{1cm} + i\oint\frac{{\rm d }w}{2\pi i}
\frac{1}{w-x_2}\log\frac{\Gamma\big[1+{i\ov
2}g\big(2\cos\varphi_1-w-{1\ov w}\big)\big]} {\Gamma\big[1-{i\ov
2}g\big(2\cos\varphi_1-w-{1\ov w}\big)\big]}\,.~~~~~~~
\nonumber\eea The function $\Phi$ is skew-symmetric
$\Phi(x_1,x_2)=-\Phi(x_2,x_1)$, and it also satisfies the
following important relation \bea\la{relP} \Phi({1\ov x_1},x_2)
+\Phi(x_1,x_2)=\Phi(0,x_2)\,,\quad |x_1|\neq 1\,, \eea which will
be used to prove the crossing equations for the dressing phase.
 If
$|x_1|=1$ this relation is modified in an obvious way due to
eq.(\ref{Phi2}).

\subsection{$\Psi$-function}

The formula (\ref{Phi2}) suggests to introduce the following
function \bea \la{Psi1} \Psi({x_1},x_2)&=&i\oint\frac{{\rm d
}w}{2\pi i} \frac{1}{w-x_2}\log\frac{\Gamma\big[1+{i\ov
2}g\big(x_1+\frac{1}{x_1}-w-{1\ov w}\big)\big]}
{\Gamma\big[1-{i\ov 2}g\big(x_1+\frac{1}{x_1}-w-{1\ov
w}\big)\big]}\,,~~~~~~~ \eea which is equal to the second term in
(\ref{Phi2}) for $x_1=e^{i\varphi_1}$. By construction,
$\Psi(x_1,x_2)=\Psi(1/x_1,x_2)$. If $x_1$ is close enough to the
unit circle then the function $\Psi({x_1},x_2)$ is well-defined
for all $|x_2|\neq 1$, and if $|x_2|=1$ we obviously can use the same
prescription (\ref{Phi1}), {\it i.e.} we approach $x_2=e^{i\vp_2}$
from exterior of the unit circle. This representation for $\Psi$
will apparently break down for such values of $x_1$ for which the
arguments of the $\Gamma$-functions in (\ref{Psi1}) become
negative integers for some $w$ on the integration contour.

\smallskip

To analyze this situation in the case $|x_2|>1$, it is convenient
to integrate  (\ref{Psi1}) by parts obtaining the following
expression \bea\la{dchi1} &&\Psi(x_1,x_2)=-{g\ov 2}\oint\frac{{\rm
d }w}{2\pi i} \log(w-x_2)\Big( 1 - {1\ov w^2}\Big)
\\\nonumber
&&~~~~~~~\times \left[\psi\Big(1+{i\ov 2}g\big(x_1+{1\ov
x_1}-w-\frac{1}{w}\big)\Big)+ \psi\Big(1-{i\ov 2}g\big(x_1+{1\ov
x_1}-w-\frac{1}{w}\big)\Big)\right]\,,~~~ \eea where the cut of
the $\log$-function should not intersect the unit
circle.\footnote{In Mathematica one could use

$\log(w-x) \to i \tan
   ^{-1}\left(\frac{w-{\rm Re}(x)}{{\rm Im}(x)}\right)+
   \frac{1}{2} \log \left(w^2-2 {\rm Re}(x)
   w+{\rm Im}(x)^2+{\rm Re}(x)^2\right)$

that gives (\ref{Psi1}) up to a constant which drops out of the
full dressing phase.} Then, instead of dealing with the cuts of
$\log\Gamma$-functions, we analyze the location of poles of the
$\psi$-functions, cf.  \cite{DHM}.

\smallskip
We use formula (\ref{dchi1}) as the definition of the function
$\Psi(x_1,x_2)$ for $|x_2|>1$ and  for all values of $x_1$ where
the integral representation is well-defined. This defines $\Psi$
as an analytic function on the $x_1$-plane with cuts. To determine
the location of the cuts, we first notice that for a generic value
of $x_1$ none of the infinitely many poles of the $\psi$-functions
in the $w$-plane falls on the unit circle. If we now start
continuously changing $x_1$, the poles start to move. At a certain
point $x_1$ might become such that  two poles of one of the
$\psi$-functions in eq.(\ref{dchi1}) reach the circle. The values
of $x_1$ for which this happens are solutions to the following
equations \bea\la{eq1}
&&x_1+{1\ov x_1}-w-\frac{1}{w} = {2i\ov g} n\,,\quad n\ge 1\,,\\
\la{eq2} &&w+\frac{1}{w}-x_1-{1\ov x_1} = {2i\ov g} n\,,\quad n\ge
1\,, \eea where $w$ is on the unit circle: $w=e^{i\theta}$.
Solutions of (\ref{eq1}) and  (\ref{eq2}) correspond to poles of
the first and second $\psi$-functions in (\ref{dchi1}),
respectively.
 It is clear that if $x_1$ and $\theta$ solve one of these equations
 then  $x_1$ and $- \theta$ also do. Thus, for each $x_1$ there are two
 values of $\theta$, and two poles of one of the $\psi$-functions are on the unit circle.

\smallskip
Equations (\ref{eq1}) and (\ref{eq2}) can be  solved in terms of
the following function $x(u)$ \bea\la{xu}
 x(u) ={u\ov 2}\left(1 +\sqrt{1-\frac{4}{u^2}}\right)\,,
\eea where the cut of $x(u)$ is from $-2$ to 2. The function
$x(u)$ maps the complex $u$-plane with the cut $[-2,2]$ onto
exterior of the unit circle, {\it i.e.} $|x(u)|>1$ for any $u$. We
are interested in solutions with $|x_1|<1$ because we will be
using the $\Psi$-function to define the dressing phase in this
region. Then the solutions to (\ref{eq1}) and (\ref{eq2}) take the
following form \bea\la{eq12} {\rm x}_\pm^{(n)}(u) = {1\ov x(u\pm
{2i\ov g} n)}\,,\quad u = w+{1\ov w}\,,\quad -2\le u\le 2\,, \eea
where the $+$ sign is for the solution to eq.(\ref{eq1}). It is
clear that all solutions to eqs.(\ref{eq1}) and
(\ref{eq2})  lie in the lower and upper half-circles,
respectively, because ${\rm Im\,}x(u +
i y)>0$ and ${\rm Im\,}x(u -
i y)<0$ for  $y> 0$.

%\vskip 0.7cm \noindent
%---------- FIGURE TOP ------------
\begin{figure}
\begin{minipage}{\textwidth}
\begin{center}
\includegraphics[width=0.3\textwidth]{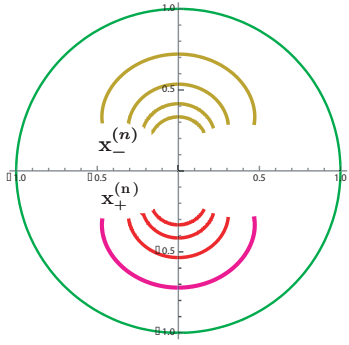}
\end{center}
\begin{center}
\parbox{5in}{\footnotesize{\caption{\label{cuts1} The curves x$_\pm^{(n)}=1/x(u\pm {2i\ov g} n)$ with $-2\le u=w+{1\ov w} \le 2$ for $g=3$
and $n=1,2,3,4$. The endpoints of the curves correspond to $w=\pm 1$. The curves closest to the circle correspond to $n=1$.}}}
\end{center}
\end{minipage}
\end{figure}
%---------- FIGURE END ------------
%\vskip -1.4cm

\smallskip
Further, we point out that eqs.(\ref{eq1}) and (\ref{eq2}) can be
thought of as the constraint equations for $x^{n\pm}$ parameters
of $n$-particle bound states, and the general solutions to these
equations can be written as \bea\la{eq1a} &&{\rm
x}_+^{(n)}=x^{n+}(z)\,,\quad w=x^{n-}(z)=e^{i\theta}\,,\\\la{eq2a}
&&{\rm x}_-^{(n)}=x^{n-}(z)\,,\quad w=x^{n+}(z)=e^{i\theta}\,.
\eea

\smallskip

The function $\Psi$ is obviously discontinuous across any curve
x$_\pm^{(n)}$. However, for $|w|=1$ the curves x$_\pm^{(n)}$ are
not closed\footnote{It is worth stressing that the unit circle
covers twice any of the curves x$_\pm^{(n)}$. } in the $x$-plane,
and, therefore, one can always reach any point inside the unit
circle without crossing them, see Figure 2. Thus, the
curves x$_\pm^{(n)}$ represent the cuts of $\Psi$. In general, one
should think of $\Psi$ as being an analytic function (for
$|x_2|>1$) on an infinite genus surface. Specifying the cut
structure as described above, defines its particular branch, that
we  will be using in this paper.

\smallskip

The jump across any of the cuts can be found by taking into
account that for any positive integer $n$
$$
\psi(z-n) = -{1\ov z} + {\rm regular~~terms}\,.
$$
Then, enclosing the poles which are approaching the unit circle,
one finds that the difference between the values of $\Psi$ on the
different edges of the cuts x$_\pm^{(n)}$ is given by
 \bea \la{Psijump1}
\lim_{\epsilon\to 0^+}\big[ \Psi(e^{\epsilon}\,{\rm
x}_+^{(n)},x_2)- \Psi(e^{-\epsilon}\,{\rm x}_+^{(n)},x_2) \big]&=&
\lim_{\epsilon\to 0^+}{ 1\ov i}\log \frac{w(e^{-\epsilon}\,{\rm
x}_+^{(n)})-x_2}{{1\ov w(e^{-\epsilon}\,{\rm x}_+^{(n)})}-x_2}
\,,\\ \la{Psijump2} \lim_{\epsilon\to 0^+}\big[
\Psi(e^{\epsilon}\,{\rm x}_-^{(n)},x_2)- \Psi(e^{-\epsilon}\,{\rm
x}_-^{(n)},x_2) \big]&=& \lim_{\epsilon\to 0^+}{ 1\ov i}\log
\frac{{1\ov w(e^{-\epsilon}\,{\rm
x}_-^{(n)})}-x_2}{w(e^{-\epsilon}\,{\rm x}_-^{(n)})-x_2} \, .
~~~~~~~ \eea Here $w(x_1)$ satisfies $|w(x_1)|<1$ and solves the
equation $x_1+{1\ov x_1}-w-\frac{1}{w} = \pm {2i\ov g}n$, where
$``+"$ sign is for eq.(\ref{Psijump1}). In deriving the formulae
above we have used that $ \lim_{\epsilon\to
0^+}w(e^{-\epsilon}\,{\rm x}_\pm^{(n)})w(e^{\epsilon}\,{\rm
x}_\pm^{(n)})= 1$ if ${\rm x}_\pm^{(n)}$ solves eqs.(\ref{eq1}),
(\ref{eq2}).

\smallskip

To treat the function $\Psi$ for  $|x_2|<1$, we use the following
identity
 \bea\la{relPs}
\Psi(x_1,x_2)=-\Psi(x_1,{1\ov x_2})+\Psi(x_1,0)\,,\quad |x_2|\neq
1\, . \eea Integrating the first term on the right hand side of this equation  by
parts, we represent it in the form (\ref{dchi1}). The second term
is defined by (\ref{Psi1}) for all values of $x_1$ except those
lying on the curves  ${\rm x}_\pm^{(n)}$ across which the function
$\Psi(x_1,0)$ is discontinuous. For a given $x_1$ we also have to
choose cuts of the integrand in such a way that they do not
intersect the unit circle. This is always possible because the
position of the branch points of the integrand coincides with the
position of the poles of the $\psi$-functions discussed above, and
for each $n$ these points come in pairs -- one pair is inside the
circle and another one is outside. For our purpose of defining the
dressing factor, it does not matter how the cuts are chosen
because different choices will lead to functions differing by an
integer multiple of $2\pi$, the latter drops out from the dressing
factor.

\smallskip

To find the jump discontinuity of $\Psi(x_1,0)$ across the cuts
${\rm x}_\pm^{(n)}$, it is convenient to first differentiate it
with respect to $x_1$ \bea\la{Psix10} &&{d\Psi(x_1,0)\ov d
x_1}=\Psi'(x_1,0)=-{g\ov 2}\Big( 1 - {1\ov
x_1^2}\Big)\oint\frac{{\rm d }w}{2\pi i} \, {1\ov w}
\\\nonumber
&&~~~~~~~\times \left[\psi\Big(1+{i\ov 2}g\big(x_1+{1\ov
x_1}-w-\frac{1}{w}\big)\Big)+ \psi\Big(1-{i\ov 2}g\big(x_1+{1\ov
x_1}-w-\frac{1}{w}\big)\Big)\right]\,.~~~ \eea Then, the
computation of
 the difference between the values of $\Psi'(x_1,0)$ on the different edges of the cuts x$_\pm^{(n)}$ follows the consideration above, and is given by
 \bea \la{Psix10jump1}
\lim_{\epsilon\to 0^+}\big[ \Psi'(e^{\epsilon}\,{\rm
x}_+^{(n)},0)- \Psi'(e^{-\epsilon}\,{\rm x}_+^{(n)},0) \big]&=&{
2\ov i} \lim_{\epsilon\to
0^+}{1-{1\ov (e^{-\epsilon}\,{\rm x}_+^{(n)})^2}\ov 1- {1\ov w(e^{-\epsilon}\,{\rm x}_+^{(n)})^2}} {1\ov w(e^{-\epsilon}\,{\rm x}_+^{(n)})} \,,\\
\la{Psix10jump2} \lim_{\epsilon\to 0^+}\big[
\Psi'(e^{\epsilon}\,{\rm x}_-^{(n)},0)- \Psi'(e^{-\epsilon}\,{\rm
x}_-^{(n)},0) \big]&=&-{ 2\ov i} \lim_{\epsilon\to 0^+}{1-{1\ov
(e^{-\epsilon}\,{\rm x}_-^{(n)})^2}\ov 1- {1\ov
w(e^{-\epsilon}\,{\rm x}_-^{(n)})^2}} {1\ov w(e^{-\epsilon}\,{\rm
x}_-^{(n)})}  \,.~~~~~~~ \eea Taking into account that \bea
{1-{1\ov x^2}\ov 1- {1\ov w(x)^2}} = {dw(x)\ov dx}\,, \eea we get
the jump discontinuity of $\Psi(x_1,0)$
 \bea \la{Psix10jump1b}
\lim_{\epsilon\to 0^+}\big[ \Psi(e^{\epsilon}\,{\rm x}_+^{(n)},0)-
\Psi(e^{-\epsilon}\,{\rm x}_+^{(n)},0) \big]&=&{ 2\ov i}
\lim_{\epsilon\to
0^+}\log w(e^{-\epsilon}\,{\rm x}_+^{(n)}) \,,\\
\la{Psix10jump2b} \lim_{\epsilon\to 0^+}\big[
\Psi(e^{\epsilon}\,{\rm x}_-^{(n)},0)- \Psi(e^{-\epsilon}\,{\rm
x}_-^{(n)},0) \big]&=&{ 2\ov i} \lim_{\epsilon\to 0^+}\log {1\ov
w(e^{-\epsilon}\,{\rm x}_-^{(n)})}  \,,~~~~~~~ \eea where the
integration constant is fixed from the requirement that the
discontinuity vanishes at the end-points of the cuts (up to an
integer multiple of $2\pi$).

\smallskip

Combining  these formulae with (\ref{Psijump1}) and
(\ref{Psijump2}), one can easily check that the jump discontinuity
of $\Psi(x_1,x_2)$ for $|x_2|<1$ is again given by the same
formulae (\ref{Psijump1}) and (\ref{Psijump2}) up to an
unimportant  integer multiple of $2\pi$.

%%%%%%%%%%%%%%%%%%%%%%%%%%%%%%%%%%%%%
\subsection{Function $\chi$ for $|x_1|\approx 1\,, |x_2|\approx 1$}

For $|x_1|>1\,, |x_2|>1$,  the DHM integral representation for the
function $\chi$ coincides with the $\Phi$-function \bea\la{c1}
\chi(x_1,x_2)&=&\Phi(x_1,x_2)\,.~~~ \eea We want to use the DHM
integral representation and the $\Phi$- and $\Psi$-functions to
fix the principal branch of the dressing phase $\theta(z_1,z_2)$,
that is to define $\theta$ for any $z_1,z_2$ on the $z$-torus. It
can be done in infinitely many ways because the phase is a
function on the direct product of two infinite-genus Riemann
surfaces. The only requirement we will impose is that for
$z_1,z_2$ being on the principal branch, the dressing phase should
satisfy the crossing equations with the function $h$ given by
(\ref{h}).

\smallskip

For the values of $x_1, x_2$ close enough to the unit circle the
analytic continuation of the function $\chi$ is in fact
unambiguous at least for $|x_{1,2}|>r_{\rm cr}$. For instance, to
determine $\chi$ for $|x_1|<1$,  we can deform the integration
contour of $w_1$, drag the point $x_1$ a little bit inside the
unit circle, and then enclose the pole at $w_1=x_1$, see Figure
3. As a result, we obtain the following
representation for $\chi$ \bea\la{cx1}
\chi({x_1},x_2)&=&\Phi(x_1,x_2)-\Psi({x_1},x_2)\,,~~~~~~~ \eea
which is valid at least in the region $1> |x_{1}|> \sqrt{1+{1\ov
g^2}}-{1\ov g}$, $|x_2|\ge 1$.

\smallskip

Similarly, for $|x_2|<1$ with $x_2$ staying close to the unit
circle, we can deform the integration contour of $w_2$ to enclose
the pole at $w_2=x_2$. This leads to the following representation
for $\chi$ \bea\la{cx2}
\chi({x_1},x_2)&=&\Phi(x_1,x_2)+\Psi({x_2},x_1)\,,~~~~~~~ \eea
where  $1>|x_{2}|> \sqrt{1+{1\ov  g^2}}-{1\ov g}$, $|x_1|\ge 1$.

%\vskip 0.7cm \noindent
%---------- FIGURE TOP ------------
\begin{figure}
\begin{minipage}{\textwidth}
\begin{center}
\includegraphics[width=0.7\textwidth]{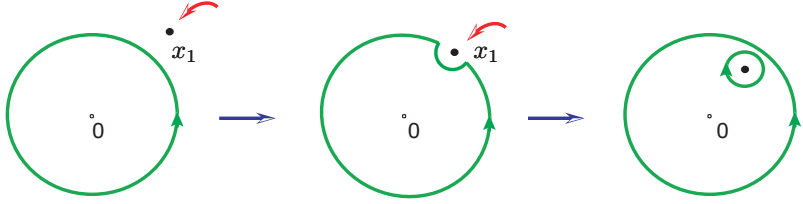}
\end{center}
\begin{center}
\parbox{5in}{\footnotesize{\caption{\label{contourBES} A little dragging
of the variable $x_1$ inside the integration contour results into
an extra contribution  given by the integral around $x_1$ with
integration performed in the clock-wise direction.
 }}}
\end{center}
\end{minipage}
\end{figure}
%---------- FIGURE END ------------
%\vskip -1.4cm

\smallskip

Finally, if both $|x_1|<1$, $|x_2|<1$ and they are close to the
unit circle, we can deform  both integration contours and
represent $\chi$ as follows \bea\nonumber
\chi(x_1,x_2)&=&\Phi(x_1,x_2)-\Psi({x_1},x_2)+\Psi({x_2},x_1)+i
\log\frac{\Gamma\big[1+{i\ov
2}g\big(x_1+\frac{1}{x_1}-x_2-\frac{1}{x_2}\big)\big]}
{\Gamma\big[1-{i\ov
2}g\big(x_1+\frac{1}{x_1}-x_2-\frac{1}{x_2}\big)\big]}\,,~~~~~~~
\eea where the last term comes from the analytic continuation of
(\ref{Psi1}) in $x_2$.

\smallskip

Since both  functions $\Phi$ and $\Psi$ are defined on the $x_1$-
and $x_2$-planes with the cuts,  the equations above define {\it a
particular} analytic continuation of  $\chi$ for all values of
$x_1$  and $x_2$. It appears, however, that this continuation of
$\chi$ is incompatible with the dressing phase considered as a
function on the $z$-torus.

\smallskip

To understand this issue,  we first notice that $x_k$ in the
functions $\chi$ appearing in the dressing phase (\ref{BES}) can
be equal to either $x^+(z_k)$ or $x^-(z_k)$. Suppose we want to
analytically continue the dressing phase $\theta(z_1,z_2)$ in the
variable $z_1$ starting from a point inside the region $
|x^\pm(z_1)|>1$, see Figure 1,  to any other point on
the $z_1$-torus. The question  is whether one could choose such a
path on the $z_1$-torus that its images $x^+(z_1)$ and $x^-(z_1)$
in the $x^+$- and $x^-$-planes would not intersect any of the
curves x$_\pm^{(n)}$. If this were possible then eq.(\ref{cx1})
provided us with a well-defined analytic continuation of $\theta$,
and the images of the curves x$_\pm^{(n)}$ on the $z$-torus would
be its cuts.  Clearly, such an analytic continuation path does not
exist only if the image of one of the curves coincides with a
one-cycle of the $z$-torus, and, therefore, it divides the torus
in two parts. In particular, from eqs.(\ref{eq1a}) and
(\ref{eq2a}) we see that for the case of fundamental particles
this happens only for curves corresponding to $n=1$ because in
this case the solutions (\ref{eq1a}) and (\ref{eq2a}) are
equivalent to conditions $|x^-|=1$ and $|x^+|=1$, respectively,
that give one-cycles of the $z$-torus, see Figure 1. In
what follows we discuss this issue in detail, and obtain analytic
continuations of the dressing phases corresponding to both the
fundamental particles and their bound states.

\section{The dressing phase for fundamental particles}\la{analyt1}
In this section we determine the analytic continuation of the
dressing phase  for fundamental particles for any $z_1, z_2$ on
the $z$-torus, and use it to prove the corresponding crossing
equations.

\subsection{Analytic continuation of the dressing phase}
The dressing factor is not a double-periodic function on the
$z$-torus, and therefore one needs to define it in the product of
two infinite strips $-{\om_1\ov 2}\le {\rm Im}(z)\le {\om_1\ov
2}$.  Since in reality we are interested in the dressing factor
$\s=e^{i\theta}$, we will not be specific about the branches of
$\log$-functions which appear in the continuation of $\chi$. We
assume here that both particles are in the fundamental
representation of the centrally extended $\su(2|2)$ superalgebra,
and the corresponding parameters $x^\pm$ obey the constraint
(\ref{const1}).

\smallskip

Each of the infinite strips is divided by the  curves $|x^\pm|=1$
into regions where $|x^\pm|$ is either greater or smaller than
unity, see Figure 1; the region with \mbox{$|x^\pm|>1$}
containing the real $z$-axis and where the dressing phase is an
analytic function of $z_1$ and $z_2$ was called the particle
region.  By shifting this region by  $\om_2$
upwards or downwards, one gets the anti-particle regions with
$|x^\pm|<1$. By shifting the real $z$-axis by ${\om_2\ov 2}$
upwards, one gets the symmetry axis of the neighboring region with
$|x^+|<1$, $|x^-|>1$ that is also the line corresponding to the
real momentum of a mirror particle.

\smallskip

In general, by shifting the real $z$-axis by an integer multiple
of ${\om_2\ov 2}$ upwards or downwards, one gets the symmetry axis
of one of these regions. Thus, it is natural to denote the
corresponding region as ${\cal R}_{n}$, and the product of two
regions as ${\cal R}_{m,n}$ where  ${\cal R}_{0,0}$ is the product
of two particle regions, and $m,n\in {\mathbb Z}$ are these
integer multiples of ${\om_2\ov 2}$ referring to the first and
second $z$-variable, respectively.

\smallskip

Since the dressing phase is antisymmetric it is sufficient to
consider only the regions
 ${\cal R}_{m,n}$ with $m\ge n$. Moreover,
 the crossing equations relate the dressing factors in the regions ${\cal R}_{m,n}$ and
 ${\cal R}_{m+2,n}$, and, therefore,
 starting from the region ${\cal R}_{0,0}$ it is enough to determine the analytic continuation of the dressing factor to the regions ${\cal R}_{1,0}$, ${\cal R}_{2,0}$, ${\cal R}_{1,1}$ and ${\cal R}_{2,1}$, and to prove the crossing equations for the regions ${\cal R}_{0,0}\leftrightarrow {\cal R}_{2,0}$ and ${\cal R}_{0,1}\leftrightarrow {\cal R}_{2,1}$.
Then,  unitarity together with crossing equations allow one to continue analytically the dressing factor to any region  ${\cal R}_{m,n}$. However, since the region  ${\cal R}_{1,1}$ contains the real momentum line of the mirror theory, we decided to analytically continue the dressing phase to the region ${\cal R}_{3,1}$ without appealing to the crossing equations, and to check  them for these regions  ${\cal R}_{1,1}\leftrightarrow {\cal R}_{3,1}$.

\subsubsection*{Region ${\cal R}_{1,0}$: $\{z_1,z_2\}\in {\cal R}_{1,0}\ \ \Longrightarrow\ \ |x_1^+|<1\,,\  |x_1^-|>1\,;\  |x_2^\pm|>1$}
First we discuss the analytic continuation from the particle
region ${\cal R}_{0,0}$ to the region ${\cal R}_{1,0}$ where
$|x_1^+|<1\,,\  |x_1^-|>1\,;\  |x_2^\pm|>1$.

\smallskip

We want to understand which curves x$_\pm^{(n)}$ are crossed when
the point $z_1$ moves upward along a vertical line in the
$z$-torus and enters the region ${\cal R}_{1,0}$, see Figure
4. In this region we need to analyze the functions
$\chi(x_1^+, x_2^\pm)$ only .
%\vskip 0.7cm \noindent
%---------- FIGURE TOP ------------
\begin{figure}
\begin{minipage}{\textwidth}
\begin{center}
\includegraphics*[width=0.27\textwidth]{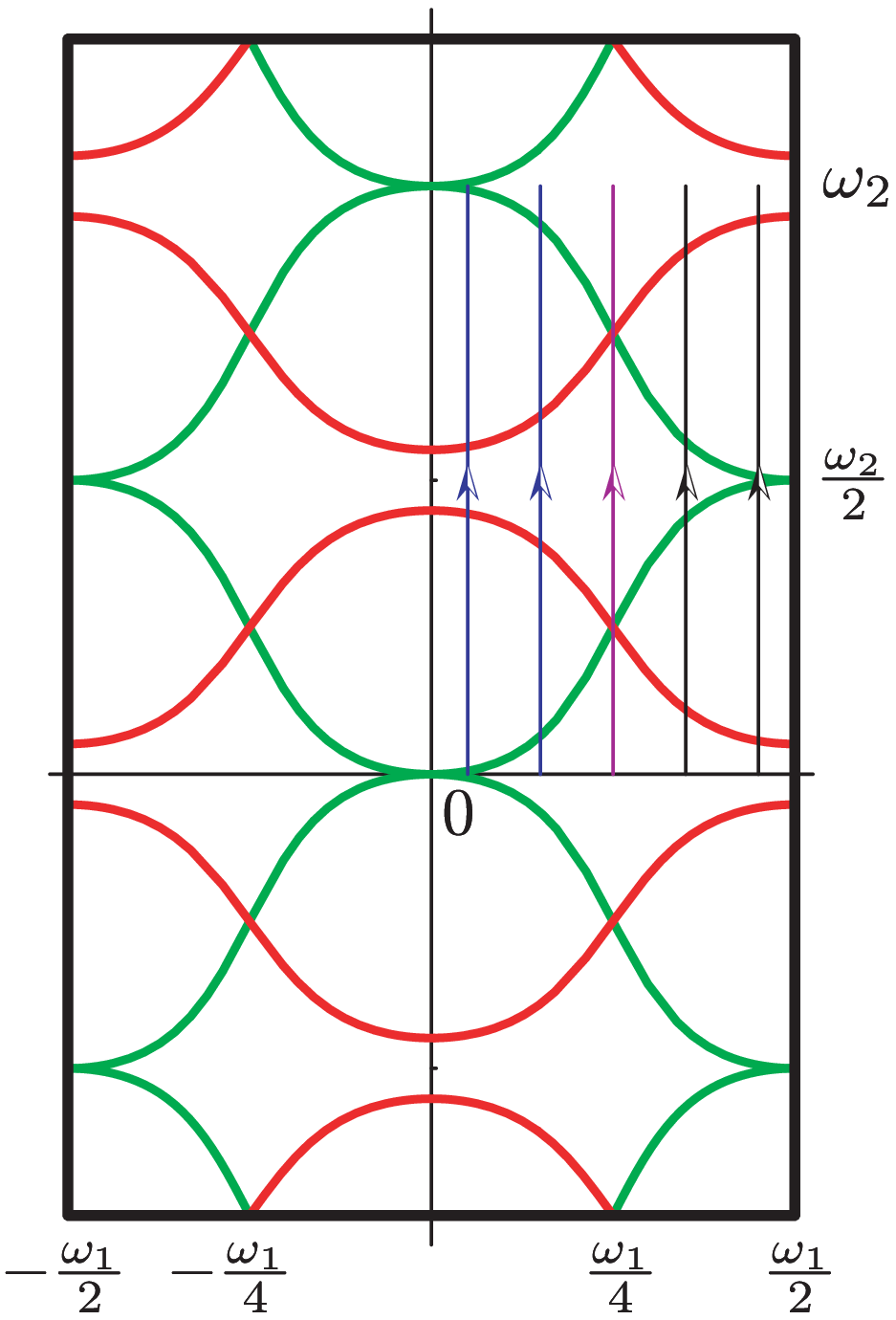}
\qquad  \includegraphics*[width=.4\textwidth]{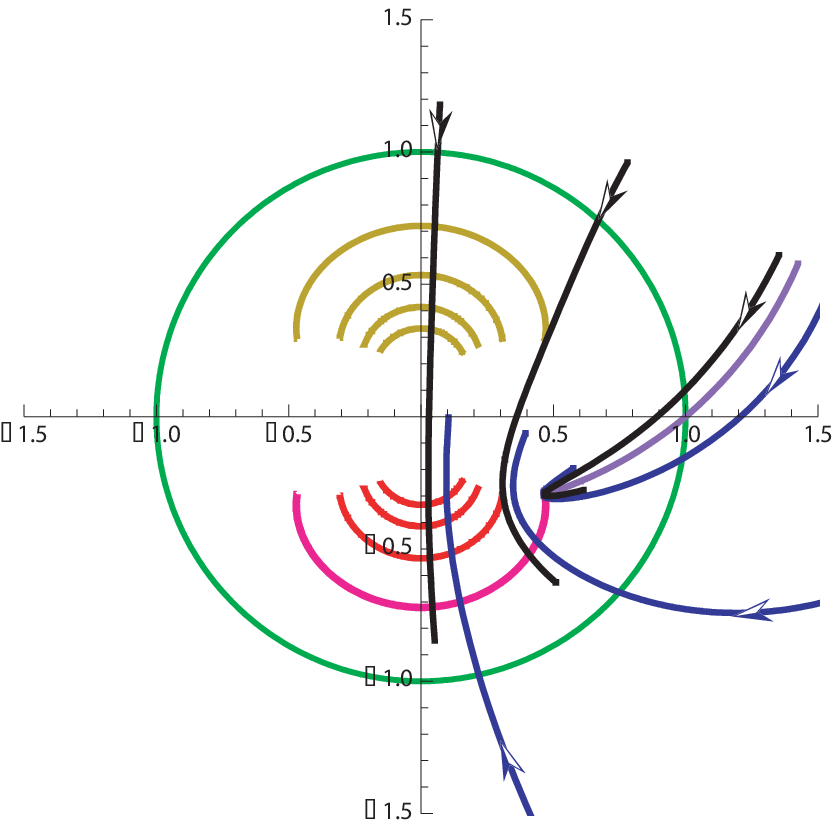}
\end{center}
\begin{center}
\parbox{5.4in}{\footnotesize{\caption{\label{torcurv} On the left figure analytic continuation  paths  are
shown on the $z$-torus. On the right figure blue curves represent
the lines $x^+(z)$ corresponding to the torus variable $z$ going
upward from the real line to the line with Im$(z)={\om_2}/i$ and
they have $|$Re$(z)|\le {\om_1\ov 4}$. Black curves $x^+(z)$
correspond to $z$ going upward and have $|$Re$(z)|\ge {\om_1\ov
4}$. Any black or blue curve intersects the lowest curve inside
the circle. Paths sufficiently close to the lines $|$Re$(z)|\le
{\om_1\ov 4}$ do not intersect any cut except the lowest curve
and, therefore, they are used for analytic continuation.}}
 }
 \end{center}
\end{minipage}
\end{figure}
%---------- FIGURE END ------------
%\vskip -1.4cm
Since the integral representation for $\chi$ is well-defined for
$|x_1^+|>r_{\rm cr}$, the dressing phase is obviously a
holomorphic function in the vicinity of the curve $|x_1^+|=1$, and
it can be evaluated there by using (\ref{cx1}). The dressing phase
cannot be however holomorphic everywhere in  ${\cal R}_{1,0}$,
because the curves  x$_\pm^{(n)}$ have images in this region, and
the function $\Psi(x_1^+,x_2^\pm)$ is discontinuous across the
curves. The images of the  curves x$_\pm^{(n)}$ in ${\cal
R}_{1,0}$ are the cuts of the dressing phase on the $z$-torus,
their end-points being the branch points of the dressing phase.

A
simple analysis reveals that no cuts are met  in the intersection
of the region $|x_1^+|<1$, $|x_1^-|>1$
 with the region Im$(x_1^\pm)<0$, see also Figure 4.
 Thus, the dressing
phase is a holomorphic function in this intersection.
The region Im$(x_1^\pm)<0$ also contains the line corresponding to the
real momentum of a mirror particle.  It was considered  as a natural candidate for the region of the mirror  theory in \cite{AFtba} because it contains one of the $2^{Q-1}$ solutions to the $Q$-particle bound state equations and is in one-to-one correspondence with the $u$-plane.  A choice of the mirror region is not however unique, and in particular
 the $Q$-particle bound state solution used in \cite{BJ}\footnote{Strictly speaking in \cite{BJ} the real momentum line
 of the mirror theory was obtained by shifting the real $z$-axis by ${\om_1\ov 2}$ downward. Obviously, these two choices are related
  to each other by reflection.} does not fall in the  region ${\rm Im}(x_1^\pm)<0$. For this reason
  we are reluctant to refer to  the region Im$(x_1^\pm)<0$ as a mirror one. Still, the region
 Im$(x_1^\pm)<0$ seems to be special because the dressing factor is analytic there.
 This follows from the consideration above, and from the fact that the dressing factor for fundamental particles is obviously analytic
 in the region ${\cal R}_{2,0}$ due to the crossing equation. We
 will return to the issue of non-uniqueness of bound state
 solutions in our conclusions.

\smallskip

Dragging $z_1$ upwards,
 we observe that the first curve the point
$x^+(z_1)$ reaches is x$_+^{(1)}=x^+(z_1)\,,\ |x^-(z_1)|=1$  that
is the lower boundary of the anti-particle region $|x_1^+|<1$,
$|x_1^-|<1$ and that is the image of the curve closest to the
circle in the lower $x^+$-half-plane.

\smallskip

We conclude, therefore, that in the case of fundamental  particles
the analytic branch of the functions $\chi(x_1^\pm, x_2^\pm)$ in
the region ${\cal R}_{1,0}$ can be defined as  \bea\la{chipap0}
{\cal R}_{1,0}:\quad \chi(x_1^+, x_2^\pm) &=& \Phi(x_1^+, x_2^\pm) - \Psi(x_1^+, x_2^\pm)\,,~~~~~~~~~~~~~~\\
\la{chimap0} \chi(x_1^-, x_2^\pm) &=& \Phi(x_1^-, x_2^\pm)\,, \eea
where $\Psi$ is given by (\ref{dchi1}).  Let us also mention that
 the region ${\cal R}_{-3,0}$  obtained by shifting the point $z_1$ downward also has
 $|x_1^+|<1$, $|x_1^-|>1$. The dressing phase, however, differs there from (\ref{chipap0})  by the double crossing term (\ref{dcross}).

\subsubsection*{Region ${\cal R}_{2,0}$:  $\{z_1,z_2\}\in {\cal R}_{2,0}\ \ \Longrightarrow\ \ |x_1^+|<1\,,\  |x_1^-|<1\,;\  |x_2^\pm|>1$}

Dragging the point $z_1$ further upward into the anti-particle
region $|x_1^\pm|<1$, we must {\it inevitably} cross the curve
x$_+^{(1)}$, the latter maps  to the one-cycle $|x_1^-|=1$ of the
$z$-torus. Then the formula (\ref{cx1}) for $\chi(x_1^+, x_2)$
should be modified because, as was discussed in the previous
section, one pole of the first $\psi$-function in  (\ref{dchi1})
moves outside the circle and another one moves inside. Therefore,
once the point $z_1$ crosses the lower boundary of the
anti-particle region $|x_1^\pm|<1$,  we should add to $\Psi$  the
following term \bea\la{dchi4} { 1\ov i}\log
\frac{w(x_1^+)-x_2^\pm}{{1\ov w(x_1^+)}-x_2^\pm}\, ,~~~ \eea where
we have taken into account the formula (\ref{Psijump1}) for the
jump discontinuity of the $\Psi$-function. Here $w(x_1^+)$ solves
the equation $x_1^++{1\ov x_1^+}-w-\frac{1}{w} = {2i\ov g}$ and
satisfies $|w(x_1^+)|<1$. Since $|x_1^-|<1$  once $z_1$ crosses
the boundary, we conclude that $w(x_1^+)=x_1^-$. Thus, we get the
following expressions for the functions $\chi$ with the first
particle being in the anti-particle region $ {\cal R}_{2,0}$
\bea\la{chipap}
{\cal R}_{2,0}:\quad \chi(x_1^+, x_2^\pm) &=& \Phi(x_1^+, x_2^\pm) - \Psi(x_1^+, x_2^\pm) +{ 1\ov i}\log \frac{{1\ov x_1^-}-x_2^\pm}{x_1^--x_2^\pm}\,,~~~~~~~~~~~~~~\\
\la{chimap} \chi(x_1^-, x_2^\pm) &=& \Phi(x_1^-, x_2^\pm) -
\Psi(x_1^-, x_2^\pm)\,,~~~~~~~~~~~~~~ \eea where $|x_1^+|<1$,
$|x_1^-|<1$ and $|x_2^\pm|>1$, and the functions $\Phi$ and $\Psi$
are given by (\ref{Phip}) and (\ref{dchi1}) for all values of
$x_1^\pm$ from the region. Let us stress that in the $z_1$-plane
the anti-particle region is obtained from the particle region
containing the real $z$-axis by shifting it by $\om_2$ upward.

\smallskip

%\vskip 0.7cm \noindent
%---------- FIGURE TOP ------------
\begin{figure}
\begin{minipage}{\textwidth}
\begin{center}
\includegraphics*[width=0.27\textwidth]{torus_curves}
\qquad
\includegraphics[width=0.45\textwidth]{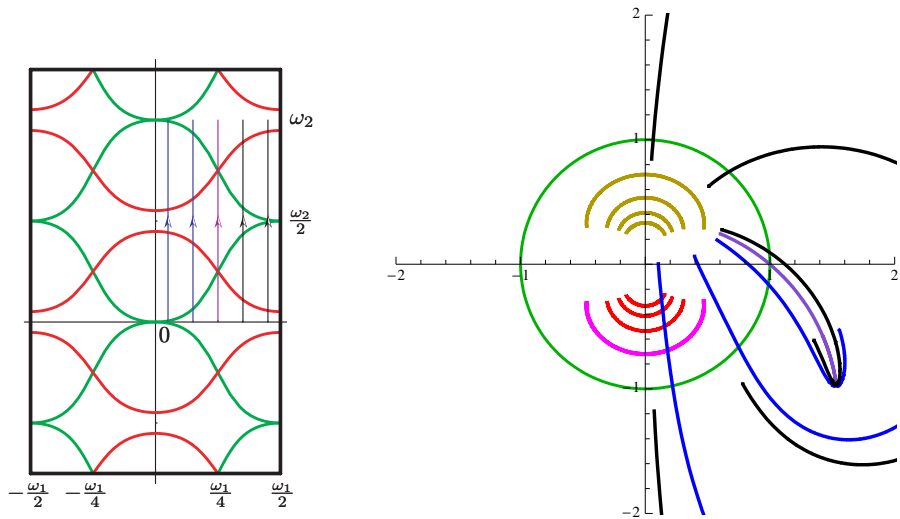}
\end{center}
\begin{center}
\parbox{5in}{\footnotesize{\caption{\label{funny_curves2} Blue and black curves on the right figure
represent the curves $x^-(z)$ corresponding to the curves $x^+(z)$
in Figure 4. No black curve intersects the cuts inside the circle,
and the blue curves  close enough to Re$(z)= {\om_1\ov 4}$ do not
intersect the cuts either.  }}}
\end{center}
\end{minipage}
\end{figure}
%---------- FIGURE END ------------
%\vskip -1.4cm

These formulae define a certain analytic branch of the functions
$\chi(x_1^\pm, x_2^\pm)$ in ${\cal R}_{2,0}$ and, as will be
discussed in the next subsection, they  are sufficient to prove
the crossing equation for $z_1,z_2$ being in the particle region.
It is worth stressing that even though the functions $\chi$ are not analytic in the region
${\cal R}_{2,0}$ because of the
cuts  located in the intersection of the anti-particle region
with the regions Im$(x^\pm)<0$ and  Im$(x^\pm)>0$, as is
evident from Figures 4 and 5,
the dressing factor itself  is analytic in ${\cal R}_{2,0}$ because it is related to the dressing factor in ${\cal R}_{0,0}$ by the crossing equation.

 We also
point out that the functions $\chi$ in (\ref{chipap}) are holomorphic
in the vicinity of the lower boundary of the anti-particle region
because these representations were derived by deforming the
integration contour in
 (\ref{Phip}) and (\ref{dchi1}). In other words, the curve
 $|x^-(z)|=1$ is not a cut of the dressing phase\footnote{It cannot
 be a cut already for the reason that it coincides with the one-cycle
 of the torus.} but rather the boundary of validity of the integral
 representation for $\chi$. Crossing this curve enforces a modification of the integral
 representation, as described above.

\smallskip

In our treatment above we have chosen a path for analytic
continuation of the dressing phase by starting with the variable
$z_1$ from the particle region and dragging it upward from the
real axes. Analogously, we could consider a path of different
orientation, {\it i.e.} the one which is obtained by shifting
$z_1$ downward from the real axis. The interested reader may
consult the appendix, where the results of this analytic
continuation are sketched. Now we find the analytic continuation
for the dressing phase in the regions ${\cal R}_{k,1}$, $k=2,3,$.
This is relevant for proving that the dressing phase of the mirror
theory also satisfies the same crossing equations.

\subsubsection*{Region ${\cal R}_{1,1}$:  $\{z_1,z_2\}\in {\cal R}_{1,1}\ \ \Longrightarrow\ \ |x_1^+|<1\,,\  |x_1^-|>1\,;\  |x_2^+|<1\,,\  |x_2^-|>1$}
Consider the case where the particles are in the region ${\cal
R}_{1,1}$ with $|x_k^+|<1$, $|x_k^-|>1$ obtained from the region
${\cal R}_{0,0}$ by moving both points $z_1$ and $z_2$ upward. If
the second particle is in the region $|x_2^\pm|>1$ the functions
$\chi$ are given by (\ref{chipap0}) and (\ref{chimap0}). If the
point $z_2$ is shifted upward, then we should add the extra
contributions coming from $\Phi$ and $\Psi$ functions, and we
derive the following expressions for the functions $\chi$
\bea\nonumber {\cal R}_{1,1}:\quad  \chi(x_1^+, x_2^+) &=&
\Phi(x_1^+, x_2^+)+ \Psi(x_2^+, x_1^+) - \Psi(x_1^+,
x_2^+)\\\nonumber &&~~~~~~~~~~~~+i \log\frac{\Gamma\big[1+{i\ov
2}g\big(x_1^++\frac{1}{x_1^+}-x_2^+-\frac{1}{x_2^+}\big)\big]}
{\Gamma\big[1-{i\ov
2}g\big(x_1^++\frac{1}{x_1^+}-x_2^+-\frac{1}{x_2^+}\big)\big]}\,,\\\nonumber
\chi(x_1^+, x_2^-) &=& \Phi(x_1^+, x_2^-) - \Psi(x_1^+, x_2^-)
\,,\\
\chi(x_1^-, x_2^+) &=& \Phi(x_1^-, x_2^+)+ \Psi(x_2^+, x_1^-)\,,\nonumber\\
\chi(x_1^-, x_2^-) &=& \Phi(x_1^-, x_2^-)\,,~~~~~~~~\la{chim0}
\eea where the last term in  the formula for $\chi(x_1^+,x_2^+)$
comes from the analytic continuation of $\Psi(x_1^+,x_2^+)$ in
$z_2$ shifted upward.

\subsubsection*{Region ${\cal R}_{2,1}$:  $\{z_1,z_2\}\in {\cal R}_{2,1}\ \ \Longrightarrow\ \ |x_1^+|<1\,,\  |x_1^-|<1\,;\  |x_2^+|<1\,,\  |x_2^-|>1$}

Shifting $z_1$ further upward into the anti-particle region, we
get \bea\nonumber {\cal R}_{2,1}:\quad  \chi(x_1^+, x_2^+) &=&
\Phi(x_1^+, x_2^+)+ \Psi(x_2^+, x_1^+) - \Psi(x_1^+, x_2^+)+{ 1\ov
i}\log \frac{{1\ov x_1^-}-x_2^+}{x_1^--x_2^+}\\\nonumber
&&~~~~~~~~~~~~~~~~+i \log\frac{\Gamma\big[1+{i\ov
2}g\big(x_1^++\frac{1}{x_1^+}-x_2^+-\frac{1}{x_2^+}\big)\big]}
{\Gamma\big[1-{i\ov
2}g\big(x_1^++\frac{1}{x_1^+}-x_2^+-\frac{1}{x_2^+}\big)\big]}\,,
\\\nonumber
\chi(x_1^+, x_2^-) &=&  \Phi(x_1^+, x_2^-)- \Psi(x_1^+, x_2^-)+{
1\ov i}\log \frac{{1\ov x_1^-}-x_2^-}{x_1^--x_2^-}\,,
\\\nonumber
\chi(x_1^-, x_2^+) &=& \Phi(x_1^-, x_2^+) - \Psi(x_1^-, x_2^+)+
\Psi(x_2^+, x_1^-)\\\nonumber &&~~~~~~~~~~~~~~~~+i
\log\frac{\Gamma\big[1+{i\ov
2}g\big(x_1^-+\frac{1}{x_1^-}-x_2^+-\frac{1}{x_2^+}\big)\big]}
{\Gamma\big[1-{i\ov
2}g\big(x_1^-+\frac{1}{x_1^-}-x_2^+-\frac{1}{x_2^+}\big)\big]}\,,
\\
\chi(x_1^-, x_2^-)&=& \Phi(x_1^-, x_2^-)- \Psi(x_1^-,
x_2^-)\,.\la{chim1} \eea

\subsubsection*{Region ${\cal R}_{3,1}$:  $\{z_1,z_2\}\in {\cal R}_{3,1}\ \ \Longrightarrow\ \ |x_1^+|>1\,,\  |x_1^-|<1\,;\  |x_2^+|<1\,,\  |x_2^-|>1$}
Dragging the point  $z_1$ further upward into the region
$|x_1^-|<1$, $|x_1^+|>1$, we cross the curve x$_-^{(1)}$ that is
mapped to the curve $|x_1^+|=1$ on the $z$-torus, the latter being
the upper boundary of the anti-particle region.

\smallskip

Then $x_1^+$ goes outside the unit circle and we need to drop
$\Psi$ function from (\ref{chipap}), and $x_1^-$ crosses
x$_-^{(1)}$, and produces an extra contribution to (\ref{chim1})
because one pole of the second $\psi$-function in  (\ref{dchi1})
moves outside the circle and another one moves inside. The result
of the analytic continuation is then given by
 \bea\nonumber {\cal R}_{3,1}:\quad
\chi(x_1^+, x_2^+) &=& \Phi(x_1^+, x_2^+)+ \Psi(x_2^+, x_1^+) +{
1\ov i}\log \frac{{1\ov x_1^-}-x_2^+}{x_1^--x_2^+}\,,
\\\nonumber
\chi(x_1^+, x_2^-) &=&  \Phi(x_1^+, x_2^-)+{ 1\ov i}\log
\frac{{1\ov x_1^-}-x_2^-}{x_1^--x_2^-}\,,
\\\nonumber
\chi(x_1^-, x_2^+) &=& \Phi(x_1^-, x_2^+) - \Psi(x_1^-, x_2^+)+
\Psi(x_2^+, x_1^-)+{ 1\ov i}\log \frac{{1\ov
x_1^+}-x_2^+}{x_1^+-x_2^+}\\\nonumber &&~~~~~~~~~~~~~~~~+i
\log\frac{\Gamma\big[1+{i\ov
2}g\big(x_1^-+\frac{1}{x_1^-}-x_2^+-\frac{1}{x_2^+}\big)\big]}
{\Gamma\big[1-{i\ov
2}g\big(x_1^-+\frac{1}{x_1^-}-x_2^+-\frac{1}{x_2^+}\big)\big]}\,,
\\
\chi(x_1^-, x_2^-)&=& \Phi(x_1^-, x_2^-)- \Psi(x_1^-, x_2^-)+{
1\ov i}\log \frac{{1\ov x_1^+}-x_2^-}{x_1^+-x_2^-}\,.\la{chir31}
\eea

Now we have all the necessary ingredients to verify the
fulfillment of the crossing equations for the dressing phase of
fundamental particles.

\subsection{The crossing equations for fundamental particles}
Having obtained the dressing phase as an analytic function with
cuts on the product of two infinite strips
$-\frac{\omega_1}{2}\leq {\rm Im}(z)\leq \frac{\omega_1}{2}$, we
can  now evaluate the left hand side of the crossing equation
$$
\Delta\theta\equiv \theta(z_1,z_2)+\theta(z_1+\om_2,z_2)\, .
$$
In terms of the $\chi$-functions $\Delta \theta$ takes the form
\bea \la{dtheta}
\begin{aligned}
 \Delta \theta &=\chi(x_1^+,x_2^+)
-\chi(x_1^+,x_2^-)-\chi(x_1^-,x_2^+)+\chi(x_1^-,x_2^-) \\
&+\chi(1/x_1^+,x_2^+)
-\chi(1/x_1^+,x_2^-)-\chi(1/x_1^-,x_2^+)+\chi(1/x_1^-,x_2^-)\, ,
\end{aligned}\eea
Considered as the function on two $z$-planes, the crossing
equation must hold for any choice of the pair $\{z_1,z_2\}$. In
what follows we will restrict ourselves to checking the crossing
equation for two different cases, namely for $\{z_1,z_2\}\in {\cal
R}_{0,0}$ and $\{z_1,z_2\}\in {\cal R}_{1,1}$. We recall that
these cases correspond to both $z_1$ and $z_2$ being in the
particle  region or in the region relevant for the mirror theory, respectively.

\smallskip

We start with the case $\{z_1,z_2\}\in {\cal R}_{0,0}\Rightarrow
|x_1^\pm|>1\,;\ |x_2^\pm|>1$. Then, $|1/x^{\pm}|<1$ and,
therefore, the arguments of the $\chi$-functions occurring in the
second line of eq.(\ref{dtheta}) are in the region ${\cal
R}_{2,0}$. Thus, evaluating the second line in eq.(\ref{dtheta}),
we have to use the formulae (\ref{chipap}) and (\ref{chimap}) with
the substitution in the latter $x_1^{\pm}\to 1/x_1^{\pm}$. Taking
into account the identity (\ref{relP}), one finds that the
contribution of $\Phi$-functions in $\Delta \theta$ cancels out
and one gets \bea \Delta\theta &=& \Psi({1\ov x_1^-},
x_2^+)-\Psi({1\ov x_1^+}, x_2^+) +\Psi({1\ov x_1^+}, x_2^-)
-\Psi({1\ov x_1^-}, x_2^-)\nonumber  \\ \la{creq1}
&&~~~~~~~~~~~~~~~~~~~~~~~~~~~~~~~~~+ {1\ov i}\log{x_1^--x_2^+\ov
{1\ov x_1^-}-x_2^+}{{1\ov x_1^-}-x_2^-\ov x_1^--x_2^-}\,. \eea The
$\Psi$-function satisfies a number of important identities which
are listed in appendix \ref{sub:app2}. In particular, by using the
formula (\ref{id1}) valid in the region $|x_1^\pm|>1$ and
$|x_2|>1$, we get \bea\nonumber \Psi({1\ov x_1^-},
x_2^+)-\Psi({1\ov x_1^+}, x_2^+) +\Psi({1\ov x_1^+}, x_2^-)
-\Psi({1\ov x_1^-}, x_2^-) &=& {1\ov i}\log{(1-{1\ov
x_1^-x_2^+})(1-{1\ov x_1^+x_2^+})\ov (1-{1\ov x_1^-x_2^-})(1-{1\ov
x_1^+x_2^-})}\, .
%\la{psi4}
\nonumber \eea Finally,
 \bea\nonumber \Delta\theta&=&{1\ov
i}\log{(1-{1\ov x_1^-x_2^+})(1-{1\ov x_1^+x_2^+})\ov (1-{1\ov
x_1^-x_2^-})(1-{1\ov x_1^+x_2^-})} +{1\ov i}\log{x_1^--x_2^+\ov
{1\ov x_1^-}-x_2^+}{{1\ov x_1^-}-x_2^-\ov x_1^--x_2^-}=
 {1\ov i} \log \Big[{x_2^-\ov x_2^+} h(x_1,x_2)\Big]\,,~~~~
\eea which is the correct crossing equation for the dressing phase
of fundamental particles.

\bigskip

Now we would like to verify the crossing equations for the case
$\{z_1,z_2\}\in {\cal R}_{1,1}\Rightarrow |x_1^+|<1\,,\
|x_1^-|>1\,;\ |x_2^+|<1\,,\  |x_2^-|>1$. Since $1/|x^+_1|>1$ and
$1/|x^-_1|<1$, the arguments of the $\chi$-functions  in the
second line of eq.(\ref{dtheta}) are in the region ${\cal
R}_{3,1}$. Thus, evaluating the second line in eq.(\ref{dtheta}),
we have to use the formulae (\ref{chim0}) and (\ref{chir31}) with
the substitution in the latter $x_1^{\pm}\to 1/x_1^{\pm}$. With
the account of  the identities (\ref{relP}) and (\ref{relPs}), we
get \bea\nonumber &&\Delta\theta = \Psi({1\ov x_1^-},
x_2^+)-\Psi({1\ov x_1^+}, x_2^+) +\Psi({1\ov x_1^+}, x_2^-)
-\Psi({1\ov x_1^-}, x_2^-) \\\la{creq11} &&~~~~~~+{1\ov
i}\log{x_1^--x_2^+\ov {1\ov x_1^-}-x_2^+}{{1\ov x_1^-}-x_2^-\ov
x_1^--x_2^-} + {1\ov i}\log{x_1^+-x_2^-\ov {1\ov
x_1^+}-x_2^-}{{1\ov x_1^+}-x_2^+\ov x_1^+-x_2^+}\\\nonumber
&&~~~~~~+{1\ov i}\log \frac{\Gamma\big[1+{i\ov
2}g\big(x_1^-+\frac{1}{x_1^-}-x_2^+-\frac{1}{x_2^+}\big)\big]}
{\Gamma\big[1-{i\ov
2}g\big(x_1^-+\frac{1}{x_1^-}-x_2^+-\frac{1}{x_2^+}\big)\big]}
\frac{\Gamma\big[1-{i\ov
2}g\big(x_1^++\frac{1}{x_1^+}-x_2^+-\frac{1}{x_2^+}\big)\big]}
{\Gamma\big[1+{i\ov
2}g\big(x_1^++\frac{1}{x_1^+}-x_2^+-\frac{1}{x_2^+}\big)\big]}\,.
\eea First, the ratio of the $\Gamma$-functions is simplified to
\bea &&{1\ov i}\log \frac{\Gamma\big[1+{i\ov
2}g\big(x_1^-+\frac{1}{x_1^-}-x_2^+-\frac{1}{x_2^+}\big)\big]}
{\Gamma\big[1-{i\ov
2}g\big(x_1^-+\frac{1}{x_1^-}-x_2^+-\frac{1}{x_2^+}\big)\big]}
\frac{\Gamma\big[1-{i\ov
2}g\big(x_1^++\frac{1}{x_1^+}-x_2^+-\frac{1}{x_2^+}\big)\big]}
{\Gamma\big[1+{i\ov
2}g\big(x_1^++\frac{1}{x_1^+}-x_2^+-\frac{1}{x_2^+}\big)\big]}~~~~\\\nonumber
&&={1\ov i}\log \frac{g^2}{4} (x_1^--x_2^+)(x_1^+ -x_2^+) (1-{1\ov
x_1^-x_2^+})(1-{1\ov x_1^+ x_2^+}) \, . \eea Second, by using the
identities (\ref{id2}) and (\ref{id5p}), one obtaines
\bea\la{id2b} \Psi({1\ov x_1^+}, x_2^-) -\Psi({1\ov x_1^-},
x_2^-)&=& {1\ov
i}\log \frac{x_2^-}{(\frac{1}{x_1^-x_2^-} -1)(x_1^+ -x_2^-)}\,, \\
\la{id4b} \Psi({1\ov x_1^-}, x_2^+) -\Psi({1\ov x_1^+},
x_2^+)&=&\frac{1}{i} \log\frac{4}{g^2}\frac{x_1^+}{x_2^+}
\frac{1}{(x_2^+-x_1^-)(x_1^+ -{1\ov x_2^+})}\, . ~~~ \eea
Substituting these results in eq.(\ref{creq11}), we again recover
the  crossing equation (\ref{cr1a}). This verification of the
crossing equation for ${\cal R}_{1,1}$ confirms correctness of our
analytic continuation of the dressing phase into this
region.

%%%%%%%%%%%%%%%%%%%%%%%%%%%%%%%%%%%%%
\section{Bound state dressing factor of string theory}\la{analyt2}
%%%%%%%%%%%%%%%%%%%%%%%%%%%%%%%%%%%%%
In this section we discuss the analytic continuation of  the
dressing phase $\theta^{QM}(z_1,z_2)$ for the scattering matrix of
$Q$-particle and $M$-particle bound states. Further, we use this
continuation to prove the general crossing equations (\ref{crMq}).

\subsection{Analytic continuation of the bound state dressing phase}

As was reviewed  in section \ref{dresscross}, in the particle
region $|x^\pm|>1$ and in terms of the variables $x^\pm$, the
dressing phase $\theta^{QM}$ has the same functional form as the
fundamental one. For this reason, it seems natural to use the
analytic continuation described in the previous section also in
the general case. It appears, however, that this continuation is
incompatible with the crossing equations (\ref{crMq}), which were
derived from the ones for fundamental particles by using the
fusion procedure.

%\vskip 0.7cm \noindent
%---------- FIGURE TOP ------------
\begin{figure}
\begin{minipage}{\textwidth}
\begin{center}
\includegraphics[width=0.4\textwidth]{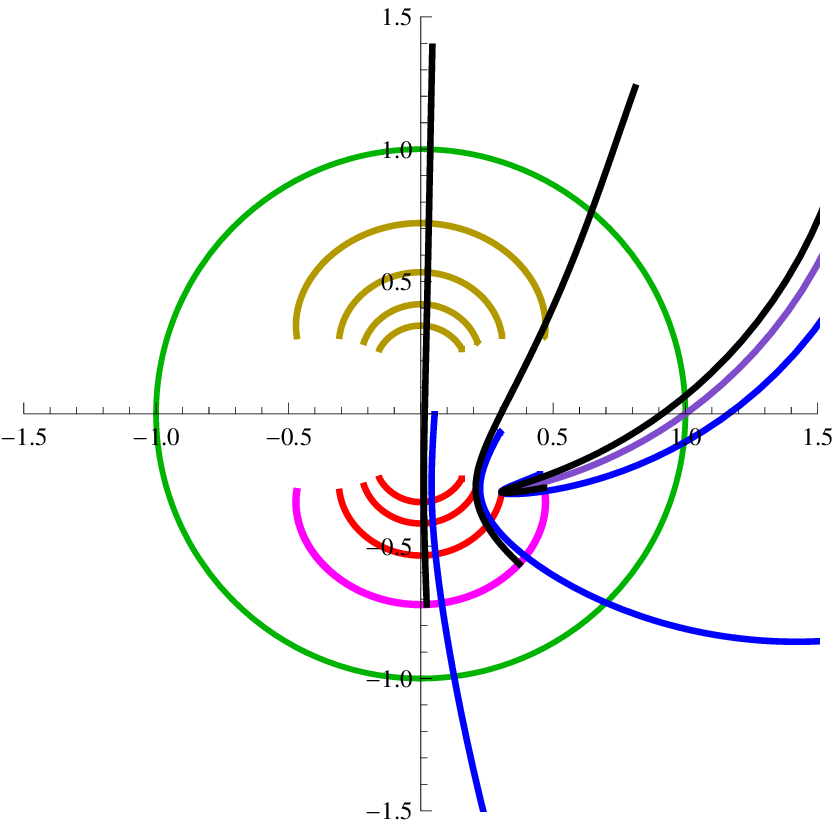}\qquad \includegraphics[width=0.4\textwidth]{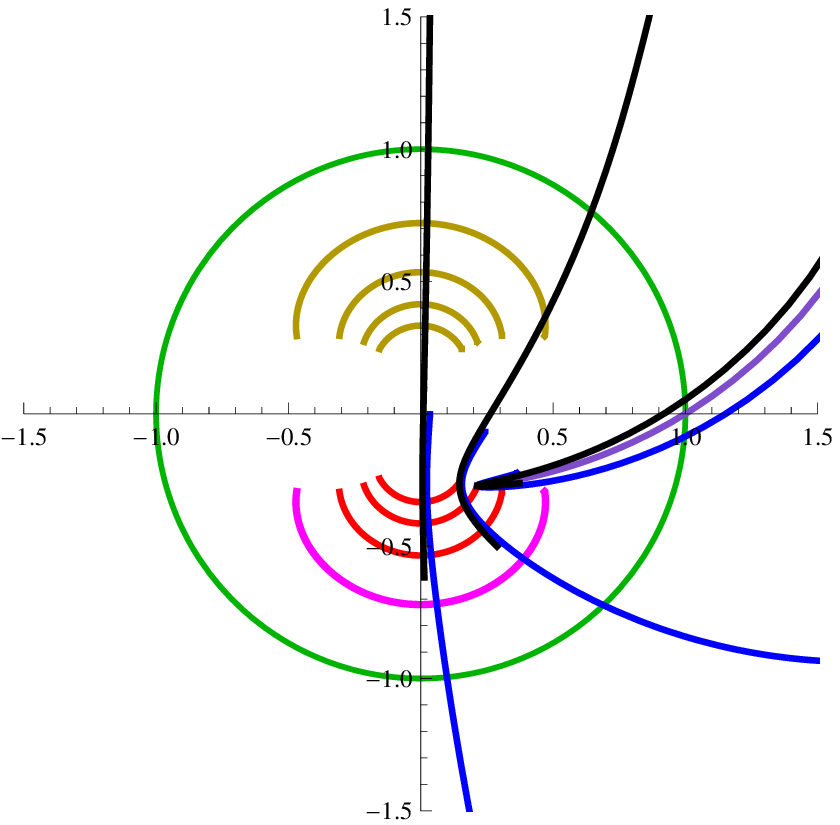}
\end{center}
\begin{center}
\parbox{5in}{\footnotesize{\caption{\label{funny_curves5} On the left picture
blue curves represent $x^+(z)$ corresponding to $z$ going upward
from the real line of the $z$-torus of a two-particle bound state
to the line with Im$(z)={\om_2}/i$ and have $|$Re$(z)|\le
{\om_1\ov 4}$. Black curves $x^+(z)$ correspond to $z$ going
upward and they have $|$Re$(z)|\ge {\om_1\ov 4}$. Any blue curve
intersects the second lower curve in the circle. On the right
picture a three-particle bound state is considered, and any blue
curve intersects the third lower curve. Curves $x^-(z)$ are shown
in Figure 10.}}}
\end{center}
\end{minipage}
\end{figure}
%---------- FIGURE END ------------
%\vskip -1.4cm

\smallskip

In this subsection we identify the analytic continuation of the
dressing phase $\theta^{QM}$ which leads
 to eqs. (\ref{crMq}). We
assume here that the first and the second particles are $Q$- and
$M$-particle bound states, respectively, and that their kinematic
parameters $x_1^\pm$ and $x_2^\pm$ obey the constraints
(\ref{xpxmq}).

\smallskip

\subsubsection*{Region ${\cal R}_{1,0}$: $\{z_1,z_2\}\in {\cal R}_{1,0}\ \ \Longrightarrow\ \ |x_1^+|<1\,,\  |x_1^-|>1\,;\  |x_2^\pm|>1$}

Again, we begin with defining the analytic continuation  from the
particle region ${\cal R}_{0,0}$ to the region ${\cal R}_{1,0}$
whose kinematic description is given above.

\smallskip

We first need to determine which curves x$_\pm^{(n)}$ are crossed
when the point $z_1$ moves upward along a vertical line in the
$z$-torus and enters the region ${\cal R}_{1,0}$, see Figure 6.
Recall that in the case of fundamental particles, the point $z_1$
in the intersection of the region ${\cal R}_{1,0}$ with the region
Im$(x^\pm)<0$ does not cross  on its way the image of any curve
x$_\pm^{(n)}$, until it reaches the curve $|x_1^-|=1$ that is the
upper boundary of ${\cal R}_{1,0}$ and an image of x$_+^{(1)}$. As
a result, the dressing phase $\theta^{11}\equiv \theta$ appears to
be a meromorphic function in this intersection. In the
$Q$-particle case the upper boundary of ${\cal R}_{1,0}$ is an
image of the curve x$_+^{(Q)}$, as is evident from eq.
(\ref{eq1a}), and there are images of the first $Q-1$ curves
x$_+^{(n)}$ in the intersection of  ${\cal R}_{1,0}$ with  the
region Im$(x^\pm)<0$. This indicates that the dressing phase
$\theta^{QM}$ should have cuts in the intersection. If one would
choose the cuts to coincide with the images of x$_+^{(n)}$, which
therefore are not allowed to be crossed, then the analytic
continuation of the dressing phase would obviously be the same as
for fundamental particles. It turns out, however, that the
crossing equations force us to analytically continue across the
curves x$_+^{(n)}$, and, therefore, the cuts in the $z$-plane
should be chosen complementary to the images of x$_+^{(n)}$.

\smallskip

A convenient and natural choice of the cuts in the $x$- and
$z$-planes  is provided by eq.(\ref{eq12}). This equation suggests
to identify the cuts with the curves $\check{\rm x}_+^{(n)}={1\ov
x(u+{2i\ov g} n)}$, $n=1,\ldots ,Q-1$, but where the parameter $u$
takes values in the region $|u|\ge 2$. In the $x$-plane all these
curves go through the origin $x=0$ corresponding to $u=\pm
\infty$, see Figure 7.
%\vskip 0.7cm \noindent
%---------- FIGURE TOP ------------
\begin{figure}
\begin{minipage}{\textwidth}
\begin{center}
\includegraphics[width=0.3\textwidth]{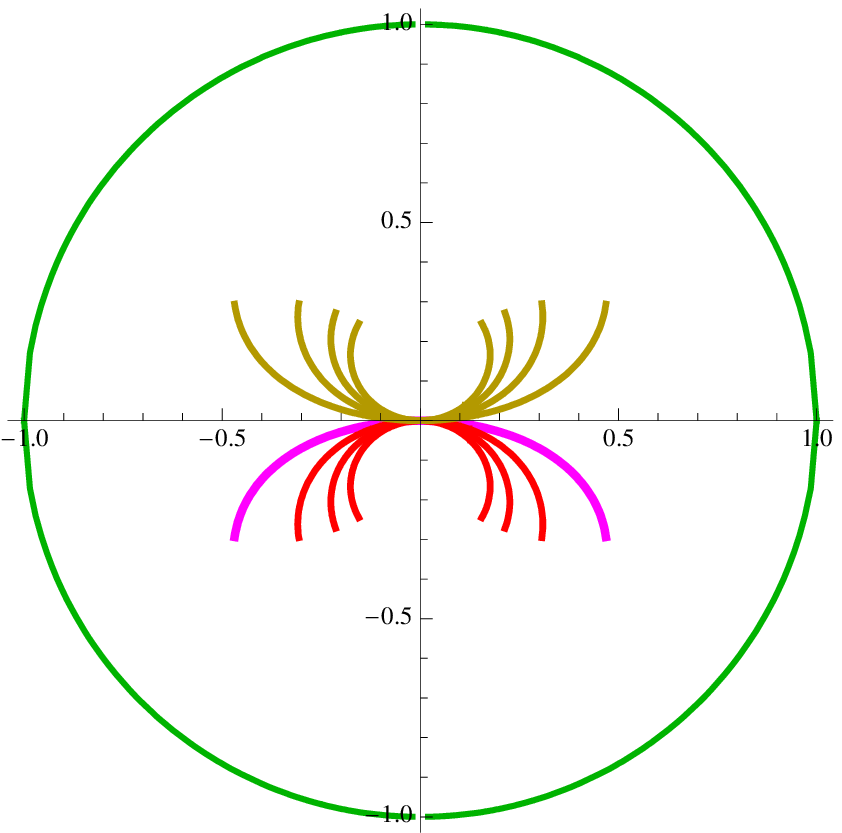} \qquad \includegraphics[width=0.3\textwidth]{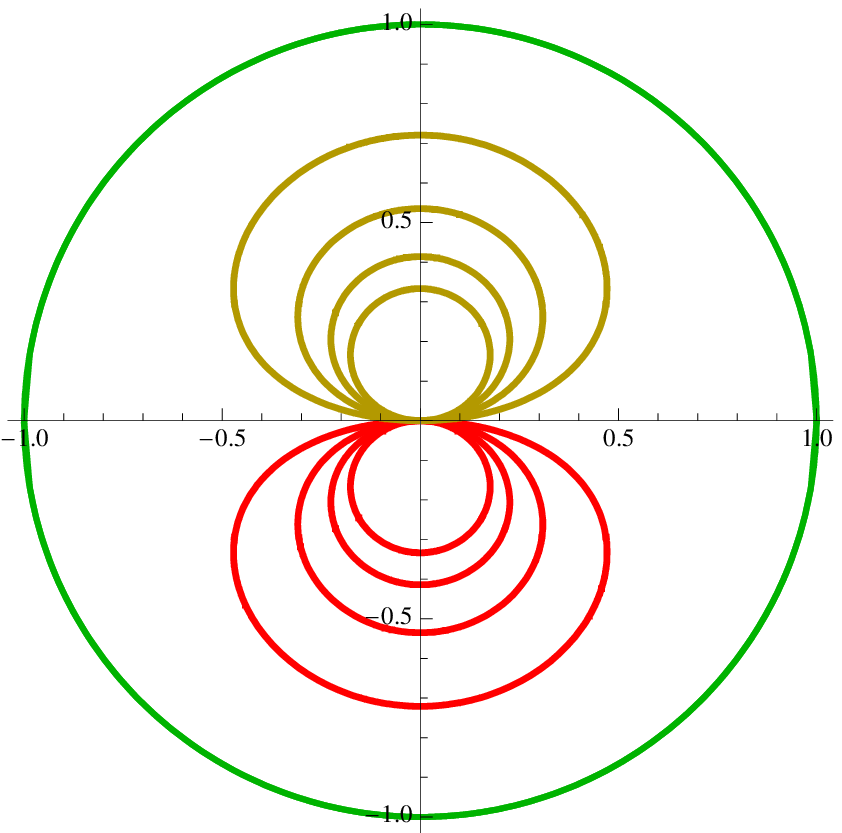}
\end{center}
\begin{center}
\parbox{5in}{\footnotesize{\caption{\label{cutsB1} On the left picture the curves $\check{\rm x}_\pm^{(n)}=1/x(u\pm {2i\ov g} n)$ with $|u| \ge 2$ for $g=3$ and $n=1,2,3,4$. The endpoints of the curves correspond to $u=\pm 2$. The curves closest to the real line correspond to $n=1$. On the right picture  the curves $1/x(u\pm {2i\ov g} n)$ with $|u| < \infty$ are shown. }}}
\end{center}
\end{minipage}
\end{figure}
%---------- FIGURE END ------------
%\vskip -1.4cm
In the $z$-plane the images of the curves are in the intersection
of the region ${\cal R}_{1,0}$ with the region
Im$(x^\pm)<0$, and the origin $x=0$ corresponds to the points
$z=-{\om_1\ov 2}+{\om_2\ov 2}$ and $z={\om_1\ov 2}+{\om_2\ov 2}$
that are one and the same point on the $z$-torus.

\smallskip

Obviously, the union of the curves ${\rm x}_+^{(n)}$ and
$\check{\rm x}_+^{(n)}$ is a closed curve in the $x$-plane, and
for $n=1,\ldots,Q-1$ its image in the region ${\cal R}_{1,0}$  is
a one-cycle and divides the $z$-torus in two parts. Let us denote
the corresponding curve in the $z$-torus as ${\cal X}_+^{(n)}$.
Thus, the region ${\cal R}_{1,0}$ is divided by these curves into
$Q$ smaller regions, see Figure 8.  We denote the region bounded
by the curves ${\cal X}_+^{(n-1)}$ and ${\cal X}_+^{(n)}$ as
${\cal R}_{1,0}^{n}$, where $n=1,\ldots,Q$. The curve ${\cal
X}_+^{(0)}$ is the lower boundary of the region ${\cal R}_{1,0}$
with $|x_1^+|=1$, while  the curve ${\cal X}_+^{(Q)} = {\rm
x}_+^{(Q)}$ is the upper boundary of ${\cal R}_{1,0}$ with
$|x_1^-|=1$.

%\vskip 0.7cm \noindent
%---------- FIGURE TOP ------------
\begin{figure}
\begin{minipage}{\textwidth}
\begin{center}
\includegraphics*[width=0.25\textwidth]{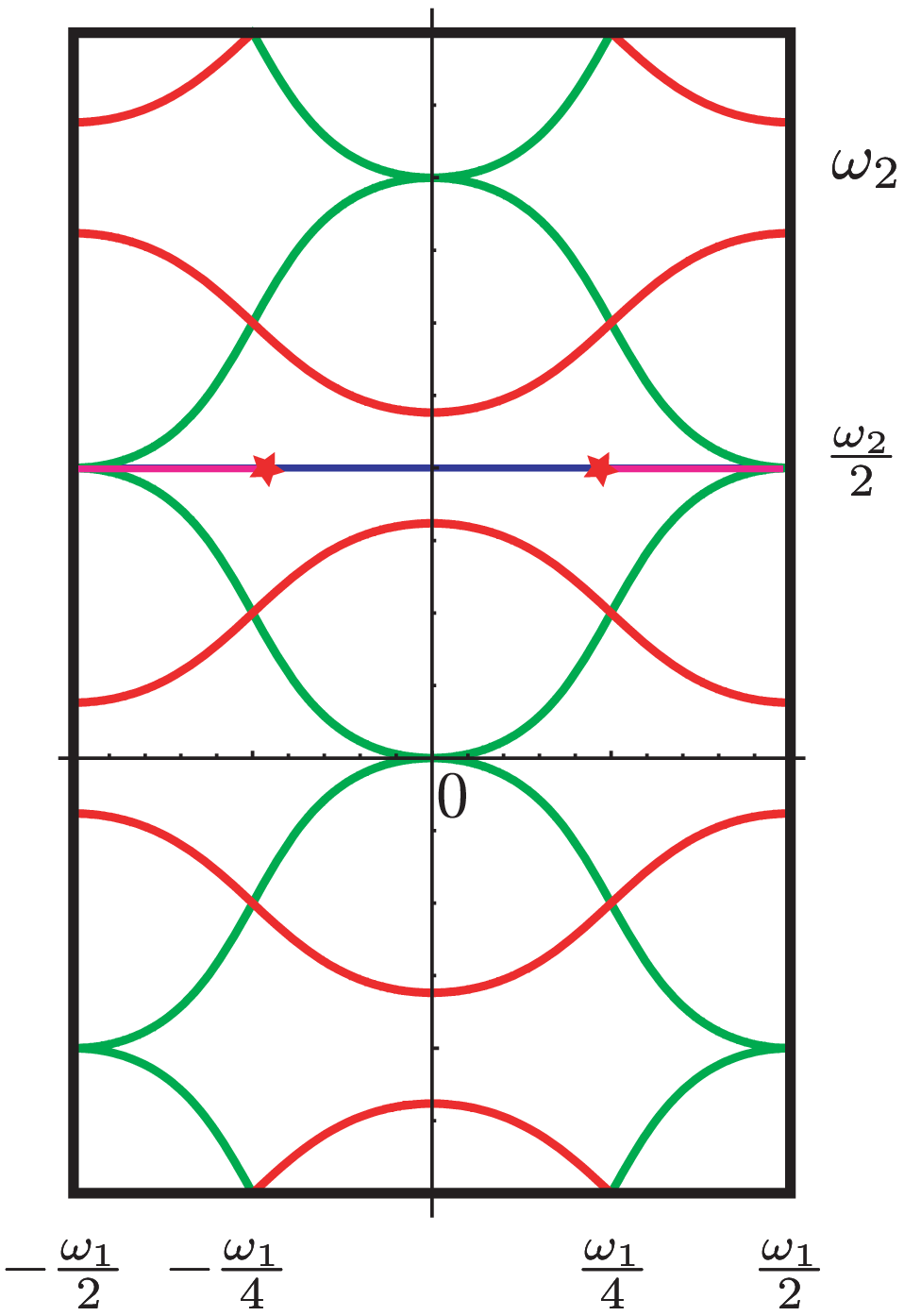}
\qquad\qquad
\includegraphics*[width=0.25\textwidth]{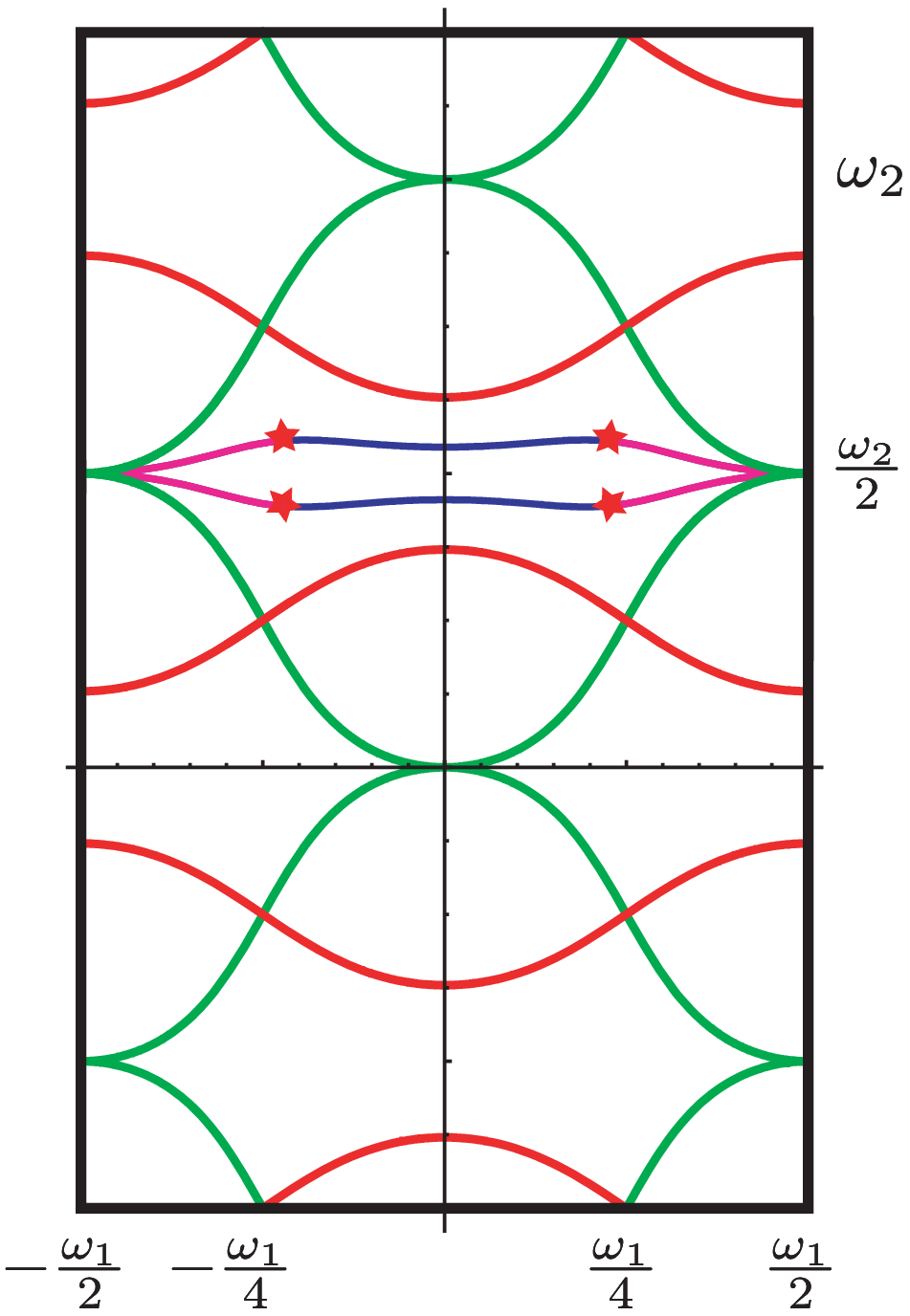}
\parbox{5in}{\caption{ \label{torus2b.eps} Division of ${\cal
R}_{1}$ into smaller regions by curves $\chi_+^{(n)}$ is shown for
two-particle (the left figure) and three-particle (the right
figure) bound states. The cuts $\check{\rm x}_+^{(n)}$ of the
dressing phase are drawn in purple. In the two-particle case, the
curve $\chi_+^{(1)}$ coincides with the real line of the mirror
theory.
 }}
\end{center}
\end{minipage}
\end{figure}
%---------- FIGURE END ------------

Thus, to reach the region ${\cal R}_{1,0}^{n}$, one should
analytically continue through the first $n-1$ curves x$_+^{(n)}$.
Therefore, in contradistinction to the case of fundamental
particles, one gets $n-1$ extra contributions. As we will see,
when properly combined, these extra contributions lead  to the
correct crossing equation.

\smallskip

We conclude, therefore, that in the  $Q$-particle bound state case
 the images of the curves $\check{\rm x}_+^{(n)}$, $n=1,\ldots,Q-1$, and x$_+^{(n)}$, $n=Q+1, \ldots \infty$, and  x$_-^{(n)}$, $n=1, \ldots \infty$ in  the region ${\cal R}_{1,0}$ are the cuts of the dressing phase on the $z$-torus,
 see Figure 9,
and
 the analytic continuation of the functions $\chi(x_1^\pm, x_2^\pm)$ in the region ${\cal R}_{1,0}^{n}$
  is given by
\bea\la{chir10b}
{\cal R}_{1,0}^{n}:\quad \chi(x_1^+, x_2^\pm) &=& \Phi(x_1^+, x_2^\pm) - \Psi(x_1^+, x_2^\pm)- { 1\ov i}\log\prod_{j=1}^{n-1} \frac{w_j^-(x_1^+)-x_2^\pm}{{1\ov w_j^-(x_1^+)}-x_2^\pm}\,,~~~~~~~~~~~~~~\\
\nonumber \chi(x_1^-, x_2^\pm) &=& \Phi(x_1^-, x_2^\pm)\,, \eea
where $n=1,\ldots, Q$. Here  we have introduced the functions
$w_j^\pm(x)$ defined as the solutions to the following equations
\bea\la{wnpm} &&x+{1\ov x}-w_j^--{1\ov w_j^-}={2i\ov g}j\,,\quad
 w_j^++{1\ov w_j^+} - x-{1\ov x}={2i\ov g}j\,,\quad |w_j^\pm|<1\,.
 \eea
 General solutions to these equations can be given in terms of the function $x(u)$ as follows
 \bea\la{gswj}
w_j^- (x_1^+)= w_{Q-j}^+(x_1^-) = {1\ov x(u_1+{i\ov g}(Q-2j))}\,,
\eea where we introduce the $u_1$-plane variable \bea u_1 = x_1^+
+ {1\ov  x_1^+} - {i\ov g}Q = x_1^- + {1\ov  x_1^-} + {i\ov g}Q\,.
\eea
In the $u_1$-plane the images of the curves $\check{\rm
x}_+^{(n)}$, $n=1,\ldots,Q-1$, and x$_+^{(n)}$, $n=Q+1, \ldots
\infty$, and  x$_-^{(n)}$, $n=1, \ldots \infty$
  are given by the following line segments
\bea\la{wnpm2}
&&\check{\rm x}_+^{(n)}:\quad u_1 =u-{i\ov g}(Q-2n)\,,\quad |u|\ge 2\,,\quad
n=1,\ldots,Q-1\,,\\
&&{\rm x}_+^{(n)}:\quad u_1 =u-{i\ov g}(Q-2n)\,,\quad |u|\le 2\,,\quad
n=Q+1,\ldots,\infty\,,\\
&&{\rm x}_-^{(n)}:\quad u_1 =u+{i\ov g}(Q-2n)\,,\quad |u|\le 2\,,\quad
n=1,\ldots,\infty\,.
 \eea

%\vskip 0.7cm \noindent
%---------- FIGURE TOP ------------
\begin{figure}
\begin{minipage}{\textwidth}
\begin{center}
\includegraphics[width=0.3\textwidth]{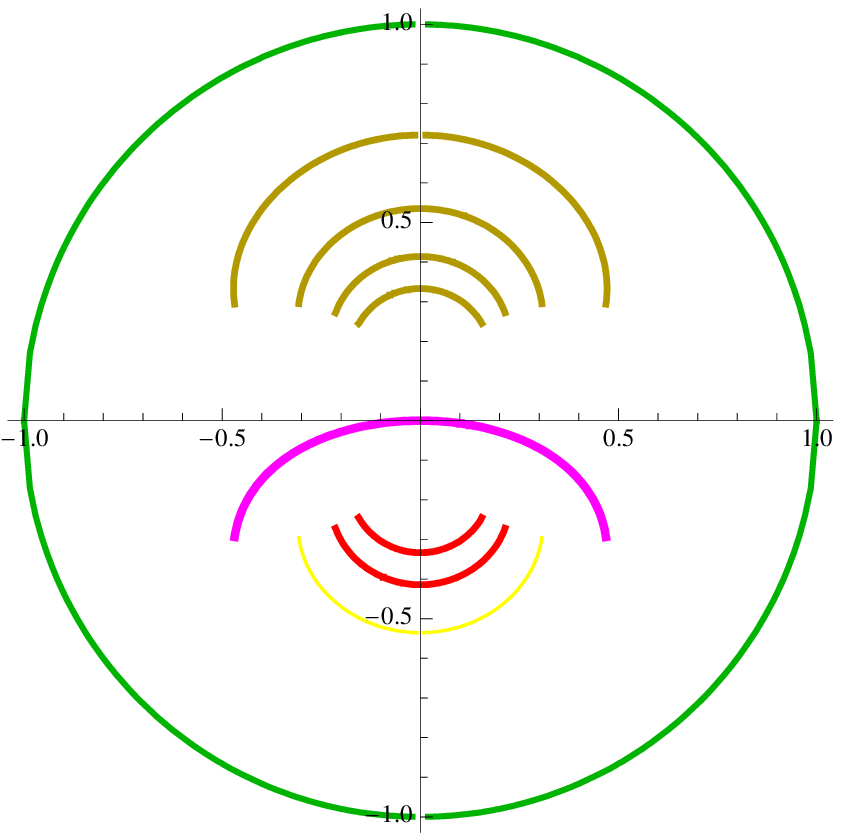}\qquad \includegraphics[width=0.3\textwidth]{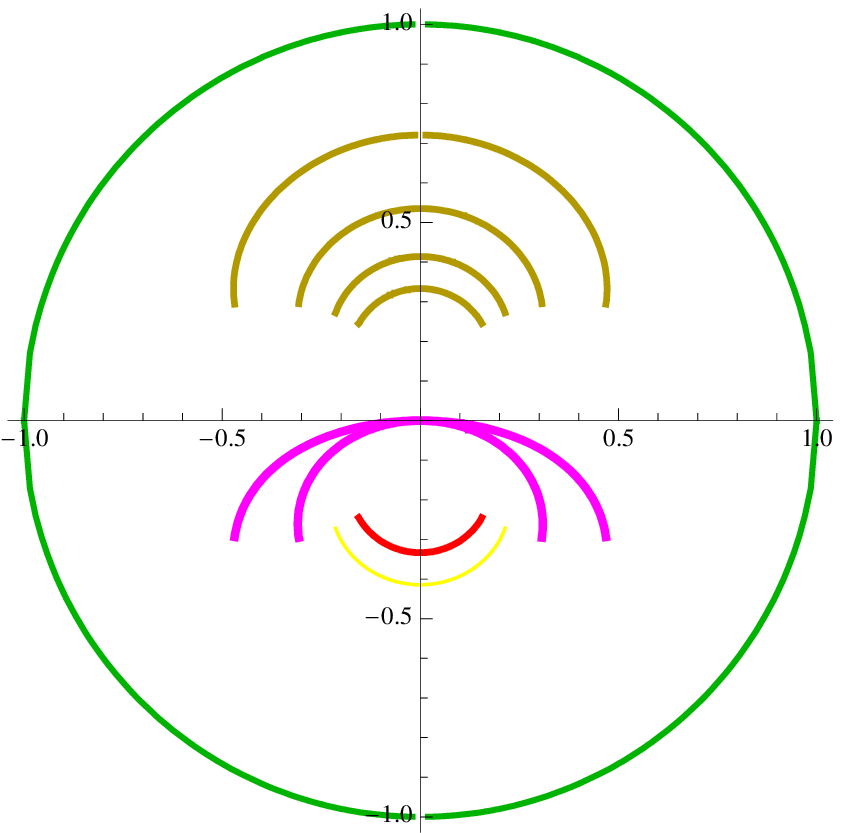}
\end{center}
\begin{center}
\parbox{5in}{\footnotesize{\caption{\label{cutsB2} On the left and right pictures the $x$-plane
the cuts of  two- and three-particle bound state dressing factors
in the region ${\cal R}_{1,0}$ are shown.  The yellow curves are
x$_+^{(Q)}$ with $Q=2,3$, and they are not cuts of the dressing
factors. }}}
\end{center}
\end{minipage}
\end{figure}
%---------- FIGURE END ------------
%\vskip -1.4cm

As was already mentioned above, the images of the curves
$\check{\rm x}_+^{(n)}$, $n=1,\ldots,Q-1$,
 and x$_+^{(n)}$, $n=Q+1, \ldots \infty$, and  x$_-^{(n)}$, $n=1, \ldots \infty$ in  the region ${\cal R}_{1,0}$
 are the cuts of the dressing phase on the $z$-torus, see Figures 8 and 9 .
We also point out that if $Q=2m$ is even then the cut with $n=m$
falls on the real line of the mirror theory! Indeed, from
eq.(\ref{wnpm2}) we see that for $\check{\rm x}_+^{(m)}$ the
parameter $u_1$ coincides with the real $u$ obeying $|u|\ge 2$.
%Under the map of the $u_1$-plane on the mirror region of the
%$z$-torus the real $u_1$-axis gets mapped onto the real line of
%the mirror region.
We also see that for $Q=2m$ the curve ${\cal
X}_+^{(m)}$ is the real line of the mirror theory.

\subsubsection*{Region ${\cal R}_{2,0}$:  $\{z_1,z_2\}\in {\cal R}_{2,0}\ \ \Longrightarrow\ \ |x_1^+|<1\,,\  |x_1^-|<1\,;\  |x_2^\pm|>1$}

Moving the point $z_1$ from the region ${\cal R}_{1,0}^{Q}$
further upward into the anti-particle region ${\cal R}_{2,0}$ with
$|x_1^\pm|<1$, we cross the curve x$_+^{(Q)}$ that is mapped to
the curve $|x_1^-|=1$ on the $z$-torus. The cut structure in this
region appears to be the same as for the fundamental particle
case, and we get the following expressions for the functions
$\chi$ with the first particle being in the region $ {\cal
R}_{2,0}$ \bea\la{chir20b} {\cal R}_{2,0}:\quad \chi(x_1^+,
x_2^\pm) &=& \Phi(x_1^+, x_2^\pm) - \Psi(x_1^+, x_2^\pm)- { 1\ov
i}\log\prod_{j=1}^{Q-1} \frac{w_j^-(x_1^+)-x_2^\pm}{{1\ov
w_j^-(x_1^+)}-x_2^\pm}\\\nonumber
 &+&{ 1\ov i}\log \frac{{1\ov x_1^-}-x_2^\pm}{x_1^--x_2^\pm}\,,~~~~~~~~~~~~~~\\
\nonumber
\chi(x_1^-, x_2^\pm) &=& \Phi(x_1^-, x_2^\pm) - \Psi(x_1^-, x_2^\pm)\,.~~~~~~~~~~~~~~
\eea
It is worth noting that the cuts of the functions $w_n^-(x_1^+)$ on the $z$-torus coincide with the images of  the curve x$_+^{(n)}$ , and, therefore, they are already included in the cut structure of the $\chi$-functions. As we will show in the next subsection, the dressing factor satisfies the crossing equation, and is an analytic function in ${\cal
R}_{2,0}$.

%\vskip 0.7cm \noindent
%---------- FIGURE TOP ------------
\begin{figure}
\begin{minipage}{\textwidth}
\begin{center}
\includegraphics[width=0.4\textwidth]{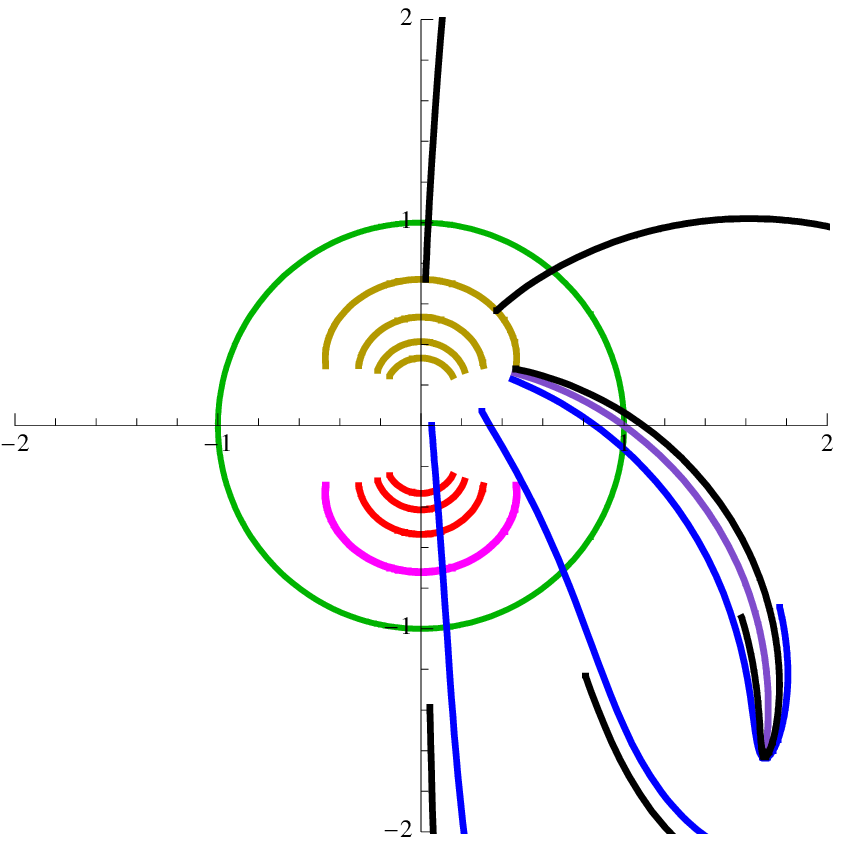}\qquad \includegraphics[width=0.4\textwidth]{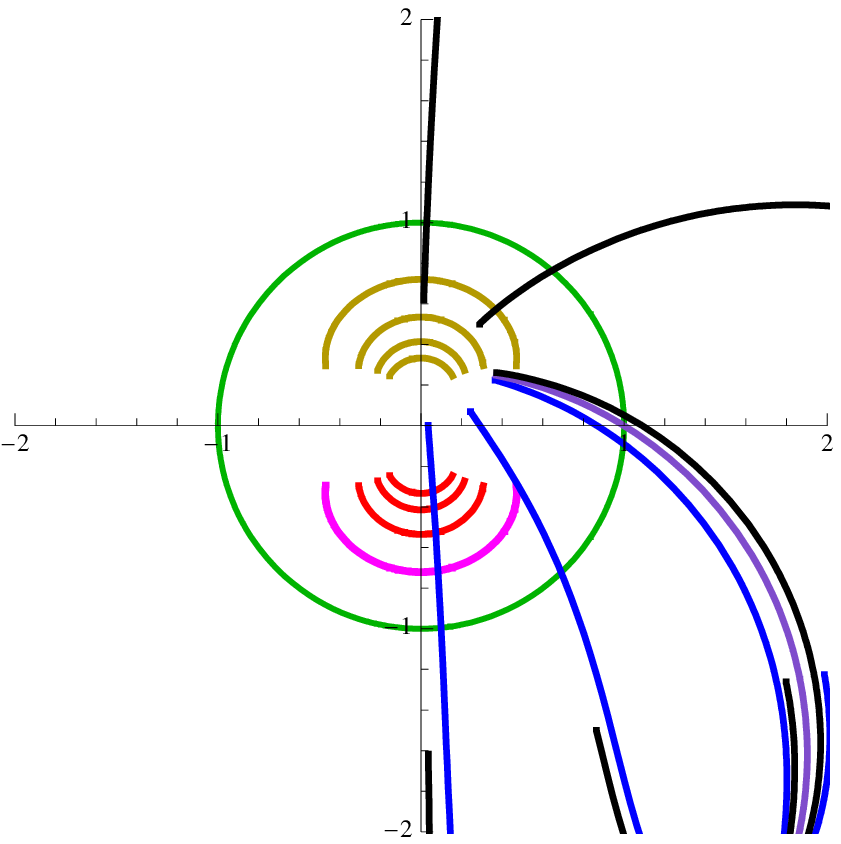}
\end{center}
\begin{center}
\parbox{5in}{\footnotesize{\caption{\label{funny_curves7}
Blue and black curves represent $x^-(z)$ corresponding to the
curves $x^+(z)$ in Figure 6. No black curve intersects the cuts
inside the circle but they all touch the upper cut. The blue
curves close enough to Re$(z)= {\om_1\ov 4}$ do not intersect the
cuts either. }}}
\end{center}
\end{minipage}
\end{figure}
%---------- FIGURE END ------------
%\vskip -1.4cm

\subsubsection*{Region ${\cal R}_{1,1}$:  $\{z_1,z_2\}\in {\cal R}_{1,1}\ \ \Longrightarrow\ \ |x_1^+|<1\,,\  |x_1^-|>1\,;\  |x_2^+|<1\,,\  |x_2^-|>1$}

Finding the analytic continuation to the region ${\cal R}_{1,1}$
basically follows the ones in the previous section, and for ${\cal
R}_{1,0}$. Since  the  second particle is an $M$-particle bound
state, the region $|x_2^+|<1$, $|x_2^-|>1$ should be divided into
$M$ smaller regions, and, as a consequence, ${\cal R}_{1,1}$
should be understood as a union of $Q\times M$ regions which we
denote as ${\cal R}_{1,1}^{n,m}$, $n=1,\ldots,Q$, $m=1,\ldots,M$.
Then, we begin with the formulae (\ref{chir10b}) for  the
functions $\chi$ in the region ${\cal R}_{1,0}$, and analytically
continue in $z_2$. The resulting expressions for the functions
$\chi$ look as follows \bea\nonumber {\cal R}_{1,1}^{n,m}:\quad &&
\chi(x_1^+, x_2^+) = \Phi(x_1^+, x_2^+)+ \Psi(x_2^+, x_1^+) -
\Psi(x_1^+, x_2^+)- { 1\ov i}\log\prod_{j=1}^{n-1}
\frac{w_j^-(x_1^+)-x_2^+}{{1\ov w_j^-(x_1^+)}-x_2^+}\\\nonumber
&&~~~~~~~~~~+ { 1\ov i}\log\prod_{j=1}^{m-1}
\frac{w_j^-(x_2^+)-x_1^+}{{1\ov w_j^-(x_2^+)}-x_1^+}+i
\log\frac{\Gamma\big[1+{i\ov
2}g\big(x_1^++\frac{1}{x_1^+}-x_2^+-\frac{1}{x_2^+}\big)\big]}
{\Gamma\big[1-{i\ov
2}g\big(x_1^++\frac{1}{x_1^+}-x_2^+-\frac{1}{x_2^+}\big)\big]}\,,\\\nonumber
&&\chi(x_1^+, x_2^-) = \Phi(x_1^+, x_2^-) - \Psi(x_1^+, x_2^-)- {
1\ov i}\log\prod_{j=1}^{n-1} \frac{w_j^-(x_1^+)-x_2^-}{{1\ov
w_j^-(x_1^+)}-x_2^-}
\,,\\
&&\chi(x_1^-, x_2^+) = \Phi(x_1^-, x_2^+)+ \Psi(x_2^+, x_1^-)+ { 1\ov i}\log\prod_{j=1}^{m-1} \frac{w_j^-(x_2^+)-x_1^-}{{1\ov w_j^-(x_2^+)}-x_1^-}\,,\nonumber\\
&&\chi(x_1^-, x_2^-) = \Phi(x_1^-, x_2^-)\,.~~~~~~~~\la{chir11b}
\eea

\subsubsection*{Region ${\cal R}_{2,1}$:  $\{z_1,z_2\}\in {\cal R}_{2,1}\ \ \Longrightarrow\ \ |x_1^+|<1\,,\  |x_1^-|<1\,;\  |x_2^+|<1\,,\  |x_2^-|>1$}

Shifting $z_1$ further upward into the anti-particle region, we
get in the region ${\cal R}_{2,1}^{~m}$, $m=1,\ldots,M$ bounded by
the curves ${\cal X}_+^{(m-1)}$ and  ${\cal X}_+^{(m)}$ in the
$z_2$-torus. The corresponding analytic continuation reads as
 \bea\nonumber {\cal R}_{2,1}^{~m}:\quad  \chi(x_1^+,
x_2^+) &=& \Phi(x_1^+, x_2^+)+ \Psi(x_2^+, x_1^+) - \Psi(x_1^+,
x_2^+)+{ 1\ov i}\log \frac{{1\ov
x_1^-}-x_2^+}{x_1^--x_2^+}\\\nonumber &&- { 1\ov
i}\log\prod_{j=1}^{Q-1} \frac{w_j^-(x_1^+)-x_2^+}{{1\ov
w_j^-(x_1^+)}-x_2^+}+ { 1\ov i}\log\prod_{j=1}^{m-1}
\frac{w_j^-(x_2^+)-x_1^+}{{1\ov w_j^-(x_2^+)}-x_1^+}\\\nonumber
&&~~~~~~~~~~~~~~~~+i \log\frac{\Gamma\big[1+{i\ov
2}g\big(x_1^++\frac{1}{x_1^+}-x_2^+-\frac{1}{x_2^+}\big)\big]}
{\Gamma\big[1-{i\ov
2}g\big(x_1^++\frac{1}{x_1^+}-x_2^+-\frac{1}{x_2^+}\big)\big]}\,,~~~~~~~~~~~~
\eea
\vspace{-0.55cm}
\bea\nonumber
~~~~~~~~~~~\chi(x_1^+, x_2^-) &=&  \Phi(x_1^+, x_2^-)- \Psi(x_1^+, x_2^-)+{ 1\ov i}\log \frac{{1\ov x_1^-}-x_2^-}{x_1^--x_2^-}- { 1\ov i}\log\prod_{j=1}^{Q-1} \frac{w_j^-(x_1^+)-x_2^-}{{1\ov w_j^-(x_1^+)}-x_2^-}\,,
\\\nonumber
\chi(x_1^-, x_2^+) &=&
\Phi(x_1^-, x_2^+) - \Psi(x_1^-, x_2^+)+ \Psi(x_2^+, x_1^-)+ { 1\ov i}\log\prod_{j=1}^{m-1} \frac{w_j^-(x_2^+)-x_1^-}{{1\ov w_j^-(x_2^+)}-x_1^-}\\\nonumber
&&~~~~~~~~~~~~~~~~+i
\log\frac{\Gamma\big[1+{i\ov
2}g\big(x_1^-+\frac{1}{x_1^-}-x_2^+-\frac{1}{x_2^+}\big)\big]}
{\Gamma\big[1-{i\ov
2}g\big(x_1^-+\frac{1}{x_1^-}-x_2^+-\frac{1}{x_2^+}\big)\big]}\,,
\\
\chi(x_1^-, x_2^-)&=& \Phi(x_1^-, x_2^-)- \Psi(x_1^-, x_2^-)\,.\la{chir21b}
\eea

\subsubsection*{Region ${\cal R}_{3,1}$:  $\{z_1,z_2\}\in {\cal R}_{3,1}\ \ \Longrightarrow\ \ |x_1^+|>1\,,\  |x_1^-|<1\,;\  |x_2^+|<1\,,\  |x_2^-|>1$}
Moving further upward into the region $|x_1^-|<1$, $|x_1^+|>1$,
  the point $z_1$  crosses the curve x$_-^{(Q)}$ that is mapped to the curve $|x_1^+|=1$ on the $z$-torus and is the upper boundary of the  region ${\cal R}_{2,1}$.
Then, the proper analytic continuation requires us to cross the images of the curves
x$_-^{(n)}$, $n=Q-1, \ldots,1$, and that means that the corresponding cuts on the $z$-torus should be the images of the curves $\check{\rm x}_-^{(n)}={1\ov x(u-{2i\ov g} n)}$, $n=Q-1, \ldots,1$, where the parameter $u$ again takes values in $|u|\ge 2$.
It is not difficult to see that for  $n=1, \ldots, Q-1$ the images of x$_-^{(n)}$ and $\check{\rm x}_-^{(n)}$ in the region ${\cal R}_{3,1}$  can be obtain from those for  x$_+^{(Q-n)}$ and $\check{\rm x}_+^{(Q-n)}$ in the region ${\cal R}_{1,1}$ just by shifting them by $\om_2$ upward.

\smallskip Thus,  the region ${\cal R}_{3,1}$ is also divided by
the curves ${\cal X}_-^{(n)}$ which are the union of the curves
${\rm x}_-^{(Q-n)}$ and $\check{\rm x}_-^{(Q-n)}$  into $Q$
smaller regions ${\cal R}_{3,1}^{n}$ bounded by the curves ${\cal
X}_-^{(n-1)}$ and ${\cal X}_-^{(n)}$, where $n=1,\ldots,Q$. To
reach the region ${\cal R}_{3,1}^{n}$, one should analytically
continue across  the  $n-1$ curves x$_-^{(Q-1)}$, $\ldots$ ,
x$_-^{(Q-n+1)}$. A new subtlety of the bound state case is that
once $x_1^-$ in the $x$-plane crosses x$_-^{(Q-n)}$, $n=1,
\ldots,Q-1$, the parameter $x_1^+$ with $|x_1^+|>1$  crosses the
curves 1/x$_-^{(n)}$ which are outside the unit circle. These
curves are the cuts of the functions $w_n^-(x_1^+)$, and,
therefore, one should replace them with $1/w_n^-(x_1^+)$ after
having crossed the cuts.

\smallskip

Repeating the consideration in the previous section and  taking
into account the new effects, one then gets that  the analytic
continuation is  given by
\bea\nonumber
{\cal R}_{3,1}^{n,m}:  &&\chi(x_1^+, x_2^+) = \Phi(x_1^+, x_2^+)+ \Psi(x_2^+, x_1^+) +{ 1\ov i}\log \frac{{1\ov x_1^-}-x_2^+}{x_1^--x_2^+}\\\nonumber
&&\hspace{-0.5cm}- { 1\ov i}\log\prod_{j=n}^{Q-1} \frac{w_j^-(x_1^+)-x_2^+}{{1\ov w_j^-(x_1^+)}-x_2^+}+ { 1\ov i}\log\prod_{j=1}^{m-1} \frac{w_j^-(x_2^+)-x_1^+}{{1\ov w_j^-(x_2^+)}-x_1^+}+ { 1\ov i}\log\prod_{j=1}^{n-1} \frac{w_j^-(x_1^+)-x_2^+}{{1\ov w_j^-(x_1^+)}-x_2^+}\,,\\
%\eea
%\vspace{-0.5cm}
%\bea
\nonumber
&&\chi(x_1^+, x_2^-) =  \Phi(x_1^+, x_2^-)+{ 1\ov i}\log \frac{{1\ov x_1^-}-x_2^-}{x_1^--x_2^-}- { 1\ov i}\log\prod_{j=n}^{Q-1} \frac{w_j^-(x_1^+)-x_2^-}{{1\ov w_j^-(x_1^+)}-x_2^-}\\\nonumber
&&~~~~~~~~~~~~~~~~~~~~~~~~~~~~~~~~~~~~~~~~~~+
{ 1\ov i}\log\prod_{j=1}^{n-1} \frac{w_j^-(x_1^+)-x_2^-}{{1\ov w_j^-(x_1^+)}-x_2^-}\,,
\\\nonumber
&&\chi(x_1^-, x_2^+) =
\Phi(x_1^-, x_2^+) - \Psi(x_1^-, x_2^+)+ \Psi(x_2^+, x_1^-)+{ 1\ov i}\log \frac{{1\ov x_1^+}-x_2^+}{x_1^+-x_2^+}
\\\nonumber
&&~~~~~~~~~~~~~~~~+ { 1\ov i}\log\prod_{j=1}^{n-1} \frac{w_{Q-j}^+(x_1^-)-x_2^+}{{1\ov w_{Q-j}^+(x_1^-)}-x_2^+}+ { 1\ov i}\log\prod_{j=1}^{m-1} \frac{w_j^-(x_2^+)-x_1^-}{{1\ov w_j^-(x_2^+)}-x_1^-}
\\\nonumber
&&~~~~~~~~~~~~~~~~+i
\log\frac{\Gamma\big[1+{i\ov
2}g\big(x_1^-+\frac{1}{x_1^-}-x_2^+-\frac{1}{x_2^+}\big)\big]}
{\Gamma\big[1-{i\ov
2}g\big(x_1^-+\frac{1}{x_1^-}-x_2^+-\frac{1}{x_2^+}\big)\big]}\,,\\
%\eea
%\vspace{-0.5cm}
%\bea
&&\chi(x_1^-, x_2^-)= \Phi(x_1^-, x_2^-)- \Psi(x_1^-, x_2^-)+{ 1\ov i}\log \frac{{1\ov x_1^+}-x_2^-}{x_1^+-x_2^-}\\\nonumber
&&~~~~~~~~~~~~~~~~~~~~~~~+ { 1\ov i}\log\prod_{j=1}^{n-1} \frac{w_{Q-j}^+(x_1^-)-x_2^-}{{1\ov w_{Q-j}^+(x_1^-)}-x_2^-}\,.\la{chir31b}
\eea
The cut structure in the region
${\cal R}_{3,1}$ is the same as in ${\cal R}_{1,1}$.
The analytic continuation of the functions $\chi$ is quite complicated. On the other hand, as we will show in the next subsection,  the dressing
phase  satisfies the crossing equation, and therefore differs from the one in the region ${\cal R}_{1,1}$ just by a simple crossing equation term.

\subsection{The crossing equations for bound states}
The crossing equations for the dressing factors  involving $Q$-particle and $M$-particle bound states are given in (\ref{crMq}), and the  $x^\pm$  variables of the  bound states  satisfy the constraints (\ref{xpxmq}). In this section we consider only the crossing equation with respect to the first argument shifted
upward from the particle region ${\cal R}_{0,0}$, and from the  region ${\cal R}_{1,1}$.

\smallskip

We start with  the particle region ${\cal R}_{0,0}$. By using
eq.(\ref{chir20b}) and the identity (\ref{relP}), we get
\bea\nonumber \Delta\theta&=& \Psi({1\ov x_1^-}, x_2^+)-\Psi({1\ov
x_1^+}, x_2^+) +\Psi({1\ov x_1^+}, x_2^-)  -\Psi({1\ov x_1^-},
x_2^-) \\\la{creq1b} &+&{1\ov i}\log{x_1^--x_2^+\ov {1\ov
x_1^-}-x_2^+}{{1\ov x_1^-}-x_2^-\ov x_1^--x_2^-} - { 1\ov
i}\log\prod_{j=1}^{Q-1} \frac{w_j^-(x_1^+)-x_2^+}{{1\ov
w_j^-(x_1^+)}-x_2^+} \frac{{1\ov
w_j^-(x_1^+)}-x_2^-}{w_j^-(x_1^+)-x_2^-}\,.\eea Taking into
account that for $|x_1^\pm|>1$ and $|x_2|>1$ \bea\la{id1b}
&&\Psi({1\ov x_1^-}, x_2) -\Psi({1\ov x_1^+}, x_2)= {1\ov
i}\oint\frac{{\rm d }w}{2\pi i} \log(w-x_2)\Big( w - {1\ov w}\Big)
\\\nonumber
&&~~~~~~~~~~~\times \sum_{j=1}^Q \left[{1\ov
(w-w_j^+(x_1^-))(w-{1\ov w_j^+(x_1^-)})}+{1\ov
(w-w_j^-(x_1^+))(w-{1\ov w_j^-(x_1^+)})}\right]~~~~~~~\\\nonumber
&&~~~~~~~=-{2Q\ov i}\log x_2 +{1\ov i}\log\prod_{j=1}^Q
(w_j^+(x_1^-)-x_2)(w_j^-(x_1^+)-x_2)\,,~~~ \eea we obtain
\bea\la{psi4b} &&\Psi({1\ov x_1^-}, x_2^+)-\Psi({1\ov x_1^+},
x_2^+) +\Psi({1\ov x_1^+}, x_2^-)  -\Psi({1\ov x_1^-}, x_2^-)
=\\\nonumber &&= {2Q\ov i}\log {x_2^-\ov x_2^+}+{1\ov i}\log{{1\ov
x_1^+} -x_2^+\ov {1\ov x_1^+} -x_2^-}{{1\ov x_1^-} -x_2^+\ov {1\ov
x_1^-} -x_2^-} +{1\ov i}\log\prod_{j=1}^{Q-1} {
(w_j^+(x_1^-)-x_2^+)(w_j^-(x_1^+)-x_2^+)\ov
(w_j^+(x_1^-)-x_2^-)(w_j^-(x_1^+)-x_2^-)}\,, \eea where we singled
out the contribution with $j=Q$. By using eq.(\ref{psi4b}), we
find that eq.(\ref{creq1b}) takes the form \bea\nonumber
\Delta\theta&=&\left(\frac{x_2^-}{x_2^+}\right)^Q \frac{x_1^-
-x_2^+}{x_1^--x_2^-}\frac{1-\frac{1}{x_1^+x_2^+}}{1-\frac{1}{x_1^+x_2^-}}
\prod_{j=1}^{Q-1} { \big(w_j^+(x_1^-)-x_2^+\big)\big(1-{1\ov
w_j^-(x_1^+)x_2^+}\big)\ov
\big(w_j^+(x_1^-)-x_2^-\big)\big(1-{1\ov
w_j^-(x_1^+)x_2^-}\big)}\,.~~~~ \eea Finally, taking into account
that $w_j^-(x_1^+) = w_{Q-j}^+(x_1^-) $, we obtain \bea\nonumber
\Delta\theta&=&\left(\frac{x_2^-}{x_2^+}\right)^Q \frac{x_1^-
-x_2^+}{x_1^--x_2^-}\frac{1-\frac{1}{x_1^+x_2^+}}{1-\frac{1}{x_1^+x_2^-}}
\prod_{j=1}^{Q-1}G(M-Q+2j)\,,~~~~ \eea that is the correct
crossing equation (\ref{crMq}).

\smallskip

To discuss the crossing equation for the region ${\cal
R}_{1,1}^{n,m}$, it is convenient to split the contribution of the
sum of phases into the following four parts \bea\la{crr11}
\Delta\theta = \Delta_1\theta +\Delta_2\theta +\Delta_3\theta
+\Delta_4\theta \,, \eea where $\Delta_1\theta$ is the
contribution of the $\Psi$-functions \bea\la{dth1} \Delta_1\theta=
\Psi({1\ov x_1^-}, x_2^+)-\Psi({1\ov x_1^+}, x_2^+) +\Psi({1\ov
x_1^+}, x_2^-)  -\Psi({1\ov x_1^-}, x_2^-)  \,, \eea
$\Delta_2\theta$ is the contribution of the $\Gamma$-functions
\bea\la{dth2} \Delta_2\theta= {1\ov i}\log
\frac{\Gamma\big[1+{i\ov
2}g\big(x_1^-+\frac{1}{x_1^-}-x_2^+-\frac{1}{x_2^+}\big)\big]}
{\Gamma\big[1-{i\ov
2}g\big(x_1^-+\frac{1}{x_1^-}-x_2^+-\frac{1}{x_2^+}\big)\big]}
\frac{\Gamma\big[1-{i\ov
2}g\big(x_1^++\frac{1}{x_1^+}-x_2^+-\frac{1}{x_2^+}\big)\big]}
{\Gamma\big[1+{i\ov
2}g\big(x_1^++\frac{1}{x_1^+}-x_2^+-\frac{1}{x_2^+}\big)\big]}\,,~~~~
\eea $\Delta_3\theta$ is the contribution due to crossing  the
upper boundaries of the regions ${\cal R}_{0,0}$, ${\cal
R}_{1,0}$, ${\cal R}_{1,1}$ and ${\cal R}_{2,1}$ \bea\la{dth3}
\Delta_3\theta={1\ov i}\log{x_1^--x_2^+\ov {1\ov
x_1^-}-x_2^+}\,{{1\ov x_1^-}-x_2^-\ov x_1^--x_2^-}\,
{x_1^+-x_2^-\ov {1\ov x_1^+}-x_2^-}\,{{1\ov x_1^+}-x_2^+\ov
x_1^+-x_2^+}\,, \eea and $\Delta_4\theta$ is the contribution due
to crossing  the curves x$_\pm^{(1)}$,..., x$_\pm^{(Q-1)}$
\bea\la{dth4} \Delta_4\theta= - { 1\ov i}\log\prod_{j=1}^{Q-1}
\frac{w_j^-(x_1^+)-x_2^+}{{1\ov w_j^-(x_1^+)}-x_2^+} \frac{{1\ov
w_j^-(x_1^+)}-x_2^-}{w_j^-(x_1^+)-x_2^-}\,.\eea We see that any
dependence on $n$ and $m$ disappears. To prove the crossing
equation, we first generalize the  identity (\ref{id5})
 \bea \la{id5g} \quad \Psi({1\ov x_1^-}, 0) -\Psi({1\ov x_1^+},
0) &=& iQ\log{g^2\ov 4}+{1\ov
i}\log\prod_{j=1}^{Q}w_j^-(x_1^+)w_j^+(x_1^-) \, ,~~~~~~~~~~~~~~
\eea Next, we take into account that in this region
$|x^+|<1$, $|x^-|>1$, and, therefore, $w_Q^-(x_1^+)= {1\ov
x_1^-}$, $w_Q^+(x_1^-)= x_2^+$. By
 using eqs.(\ref{id1b}), (\ref{relPs}) and (\ref{id5g}), we find
\bea\nonumber
&&\Delta_1\theta= {Q\ov i}\log {4\ov g^2} + {2Q-1\ov i}\log {x_2^-\ov x_2^+} -
{1\ov i}\log (x_1^+-x_2^-)  (x_1^--x_2^+) (1-{1\ov x_1^+x_2^+}) (1-{1\ov x_1^-x_2^-}) \\
\la{dth1b}
&&~~-{1\ov i}\log\prod_{j=1}^{Q-1}(w_j^-(x_1^+)-x_2^-)(1-{1\ov w_j^-(x_1^+)x_2^+})(w_j^+(x_1^-)-x_2^-)(1-{1\ov w_j^+(x_1^-)x_2^+})
\,.~~~
\eea
Simplifying $\Delta_2\theta$, we get
\bea\nonumber
&&\Delta_2\theta={Q\ov i}\log {g^2 \ov 4} +
{1\ov i}\log (x_1^+-x_2^+)  (x_1^--x_2^+) (1-{1\ov x_1^-x_2^+}) (1-{1\ov x_1^+x_2^+}) \\
\la{dth2b} &&+{1\ov
i}\log\prod_{j=1}^{Q-1}(w_j^-(x_1^+)-x_2^+)(1-{1\ov
w_j^-(x_1^+)x_2^+})(w_j^+(x_1^-)-x_2^+)(1-{1\ov
w_j^+(x_1^-)x_2^+}) \, .~~~~~~~\eea Summing up all the
contributions, we again obtain the crossing equation (\ref{crMq}).

\section{Bound state dressing factor of mirror theory}
In this section we determine the dressing factor which encodes
scattering of bound states in the mirror theory and analyze some
of its properties. Recall that $Q$-particle bound states of the
mirror theory obey the following equations \bea\la{bse}
x_1^-=x_2^+\,,\ \ x_2^-=x_3^+\,,\ldots, \ \ x_{Q-1}^-=x_Q^+\, .\
\eea These equations have $2^{Q-1}$ different solutions sharing
the same set of conserved charges. One of these solutions has all
the constituent particles located in the region  Im$(x_j^\pm)<0$,
which has been identified in \cite{AFtba} as the mirror region. In
general, the constituent particles lie anywhere on the $z$-torus,
and, therefore, their individual dressing factors should be
determined by using the analytic continuation we have established
in section 4.

\smallskip

As in the string theory case, the mirror bound state dressing
factor $\sigma^{QQ'}$ can be found by fusing the dressing factors
of the constituent particles. However, in dealing with TBA
equations, we are primarily interested not in the dressing factor
itself but rather in the following quantity \bea \Sigma^{QQ'} =
\sigma^{QQ'}\,\prod_{j=1}^Q\prod_{k=1}^{Q'} {1-{1\ov x_j^+
z_k^-}\ov 1-{1\ov x^-_j z^+_k}}\, ,\la{Sf}\eea because its
logarithmic derivative appears as one of the TBA kernels
\cite{AFmtba}. Here $x_j^{\pm}$ and $z_k^{\pm}$ are the
kinematical parameters of the constituent particles corresponding
to $Q$- and $Q'$-particle bound states, respectively, and \bea
\sigma^{QQ'}=\,\prod_{j=1}^Q\prod_{k=1}^{Q'}\sigma(x_j,z_k)\, .
\eea
 The quantity
$\Sigma^{QQ'}$ arises from fusion of the scalar factors of mirror
scattering matrices corresponding to the constituent particles.
Since the bound state equations have many solutions, it is a
priori unclear which one should be used in the fusion procedure.
Also, in contrast to the string theory bound state S-matrix, the
product factor on the right hand side of eq.(\ref{Sf}) does depend
on the internal structure of the bound states involved. Indeed,
defining the bound state kinematic parameters as \bea
y_1^+=x_1^+\,,~~~ y_1^-=x_Q^-\,,~~~y_2^+=z_1^+\,,~~~
y_2^-=z_{Q'}^-\,  \eea and using the bound state equations, the
product factor can be represented as \bea\la{Dthtot}
\prod_{k=1}^{Q'}\prod_{j=1}^Q { 1- {1\ov x_j^+z_k^-}\ov 1-{1\ov
x_j^-z_k^+}} =
 { 1- {1\ov y_1^+y_2^-}\ov 1-{1\ov y_1^-y_2^+}}\prod_{j=1}^{Q-1} { 1- {1\ov x_j^-y_2^-}\ov 1-{1\ov x_j^-y_2^+}}
 \prod_{k=1}^{Q'-1}{ 1- {1\ov y_1^+z_k^-}\ov 1-{1\ov
 y_1^-z_k^-}}\, ,
\eea which makes this dependence manifest. On the other hand, in
the physical mirror theory we might expect to find a unique bound
state scattering matrix. Indeed, as we will argue, the choice of a
bound state solution is just a matter of convenience and all
$2^{Q-1}$ solutions lead to one and the same scalar factor
$\Sigma^{QQ'}$ and, as a result, to the same bound state S-matrix.

\smallskip

Before studying the dressing factor in full generality,  we
consider $\Sigma^{Q1}$, and evaluate it on a particular bound state
solution. The most convenient choice is provided by the bound
state solution used in \cite{BJ}. Indeed, for this solution only
the first particle occurs in the region ${\cal R}_{1}$, while
all the others fall in the particle region: \bea\la{mr2}
 |x_j^-| >1\,,\ |x_{j+1}^+| >1\,,\ j=1,\ldots,Q\,,\quad |x_1^+| <1\,,\ {\rm Im}(x_1^+) <0\,,\ {\rm Im}(x_Q^-) <0\, ~~~~~~
\eea and, therefore, this solution requires a minimal amount of
analytic continuation. In terms of the function $x(u)$ given by
(\ref{xu}) this solution reads as \bea\la{sol2}
&&x_j^-=x\big(u+{i\ov g}(Q-2j)\big)\,,\quad j=1,\ldots,Q\,,
~~~~\\\nonumber &&x_1^+={1\ov x\big(u+{i\ov g}Q\big)}\,,\quad
x_j^+=x\big(u+{i\ov g}(Q-2j+2)\big)\,,\quad j=2,\ldots,Q\,.~~~~
\eea For $Q=2m$  the middle particles with $j=m, m+1$ can be on
the cut of $x(u)$.

\smallskip

As a warm up exercise, we will first consider the case where  the
second particle is a fundamental particle located in the particle
region. The dressing phase is obtained through the fusion
procedure \bea\la{thfus} \theta(y_1,y_2) =
\sum_{j=1}^Q\theta(x_j,y_2) \,, \quad y_1^+=x_1^+\,,\quad
y_1^-=x_Q^-\,, \eea where $|y_2^\pm|>1$ since these parameters
correspond to the second particle. A simple computation making use
of solution (\ref{sol2}) gives \bea\la{thQ1c} {1\ov
i}\log\Sigma^{Q1}(y_1,y_2) &=&
 \Phi(y_1^+,y_2^+) - \Psi(y_1^+,y_2^+) - \Phi(y_1^+,y_2^-)+ \Psi(y_1^+,y_2^-)\\\nonumber
&-&\Phi(y_1^-,y_2^+) + \Phi(y_1^-,y_2^-)+ {1\ov i}\log{ 1- {1\ov
y_1^+y_2^-}\ov 1-{1\ov y_1^-y_2^+}}\prod_{j=1}^{Q-1} { 1- {1\ov
x\big(u+{i\ov g}(Q-2j)\big)y_2^-}\ov 1-{1\ov x\big(u+{i\ov
g}(Q-2j)\big)y_2^+}}\,. \eea Here we have also applied
 the
formulae (\ref{chipap0}) and (\ref{chimap0}) to account for the
dressing phase of the first particle being in the region ${\cal
R}_{1}$; the logarithmic factor on the right hand side comes from
eq.(\ref{Dthtot}).

\smallskip

The logarithmic term in the expression (\ref{thQ1c}) exhibits an
explicit dependence on the kinematic parameters of the constituent
particles. As we will now see,  this dependence is, however,
artificial and can be completely removed by making appropriate
transformations of $\Psi$-functions. First, as is shown in
appendix \ref{sub:app4}, the following formula is valid
 \bea\la{id1b3} \hspace{-0.8cm}
\Psi(y_{1}^+, y_2^-) -\Psi(y_{1}^-, y_2^-)= -{1\ov
i}\log\left({y_1^+\ov y_2^-}-1 \right)\left({1\ov y_1^- y_2^-}-1
\right)\prod_{j=1}^{Q-1}\left(1-{1\ov x_j^-y_2^-}\right)^2. \eea
Second, we rewrite eq.(\ref{thQ1c}) in the following form
\bea\la{thQ1cc} {1\ov i}\log\Sigma^{Q1}(y_1,y_2) &=&
 \Phi(y_1^+,y_2^+)- \Phi(y_1^+,y_2^-)-\Phi(y_1^-,y_2^+)+
 \Phi(y_1^-,y_2^-)\\
&+&\frac{1}{2}\Big[ - \Psi(y_1^+,y_2^+) -
\underbrace{\Psi(y_1^+,y_2^+)}+ \Psi(y_1^+,y_2^-)+
\underbrace{\Psi(y_1^+,y_2^-)}\Big] \nonumber
 \\\nonumber
&&~~~~~~~~~~~~~~~~~~~~~~+ {1\ov i}\log{ 1- {1\ov y_1^+y_2^-}\ov
1-{1\ov y_1^-y_2^+}}\prod_{j=1}^{Q-1} { 1- {1\ov x\big(u+{i\ov
g}(Q-2j)\big)y_2^-}\ov 1-{1\ov x\big(u+{i\ov
g}(Q-2j)\big)y_2^+}}\,. \eea
Finally, by using the formula
(\ref{id1b3}), we substitute the first and the second underbraced
$\Psi$-functions for $\Psi(y_1^-,y_2^+)$ and $\Psi(y_1^-,y_2^-)$,
respectively, and obtain the following neat result
\bea \la{resQ1}
{1\ov i}\log\Sigma^{Q1}(y_1,y_2) &=&
 \Phi(y_1^+,y_2^+)- \Phi(y_1^+,y_2^-)-\Phi(y_1^-,y_2^+)+
 \Phi(y_1^-,y_2^-)\\
&+&\frac{1}{2}\Big[ - \Psi(y_1^+,y_2^+) - \Psi(y_1^-,y_2^+)+
\Psi(y_1^+,y_2^-)+ \Psi(y_1^-,y_2^-)\Big] \nonumber\\
&&~~~~~~~~~~~~~~~~~~~~
+\frac{1}{2i}\log\frac{(y_1^+-y_2^+)\Big(y_2^-
-\frac{1}{y_1^+}\Big)^2}{(y_1^+-y_2^-)\Big(y_2^-
-\frac{1}{y_1^-}\Big)\Big(y_2^+
-\frac{1}{y_1^-}\Big)}\, .
 \nonumber \eea
Quite fascinating, all the dependence on the constituent particles
has completely cancelled out! Although, to obtain the dressing
factor we started from a particular bound state solution, the
final result does not bear any reminiscence of this particularity.
This provides a strong indication that  the same universal answer
will be obtained by starting from any of $2^{Q-1}$ solutions. In
fact, in appendix \ref{sub:app3} we provide another derivation of
the corresponding dressing factor starting from the bound state
solution with all constituent particles being in the mirror region
${\rm Im}\, x^{\pm}<0$, and show that it leads to the same  answer
as above.

\smallskip

Now we are ready to obtain the general bound state factor
$\Sigma^{QQ'}$ of the mirror theory. Our success in showing the
decoupling of bound state constituents from the final answer
motivates us to start again from the solution (\ref{sol2}) for
both $Q$ and $Q'$ bound states. Applying a similar reasoning as
before, we find \bea\la{sigtot2}
\begin{aligned}{1\ov i}\log\Sigma^{QQ'}(y_1,y_2) &=
\Phi(y_1^+,y_2^+)-\Phi(y_1^+,y_2^-)-\Phi(y_1^-,y_2^+)+\Phi(y_1^-,y_2^-)
\\ &-
\Psi(y_1^+,y_2^+)+\Psi(y_1^+,y_2^-)+\Psi(y_{2}^+,y_1^+)-\Psi(y_{2}^+,y_1^-)
\\ &+{1\ov i}\log \frac{\Gamma\big[1-{i\ov
2}g\big(y_1^++\frac{1}{y_1^+}-y_2^+-\frac{1}{y_2^+}\big)\big]}
{\Gamma\big[1+{i\ov
2}g\big(y_1^++\frac{1}{y_1^+}-y_2^+-\frac{1}{y_2^+}\big)\big]}
\\
&+ {1\ov i}\log {1- {1\ov y_1^+y_2^-}\ov 1-{1\ov
y_1^-y_2^+}}\prod_{j=1}^{Q-1} { 1- {1\ov x_j^-y_2^-}\ov 1-{1\ov
x_j^-y_2^+}} \prod_{k=1}^{Q'-1}{ 1- {1\ov y_1^+z_k^-}\ov 1-{1\ov
y_1^-z_k^-}}\,. \end{aligned} \eea This formula was derived by
summing up the corresponding contributions coming from four
regions ${\cal R}_{1,1}$, ${\cal R}_{1,0}$, ${\cal R}_{0,1}$ and
${\cal R}_{0,0}$. The logarithmic factor in the last line comes
from eq.(\ref{Dthtot}) and, as before, it contains the explicit
dependence on the bound state constituents.

\smallskip

To get rid of the bound state constituents in eq.(\ref{sigtot2}),
we can try a similar trick as in the previous case. By using the
formula (\ref{prima!})  worked out in appendix \ref{sub:app4}, we
find
 \bea\la{sigtot3} \begin{aligned} {1\ov i}\log\Sigma^{QQ'}(y_1,y_2)
 &=
\Phi(y_1^+,y_2^+)-\Phi(y_1^+,y_2^-)-\Phi(y_1^-,y_2^+)+\Phi(y_1^-,y_2^-)
\\ &-{1\ov 2}\left(\Psi(y_1^+,y_2^+)+\Psi(y_1^-,y_2^+)-\Psi(y_1^+,y_2^-)-\Psi(y_1^-,y_2^-)\right) \\
&+{1\ov
2}\left(\Psi(y_{2}^+,y_1^+)+\Psi(y_{2}^-,y_1^+)-\Psi(y_{2}^+,y_1^-)
-\Psi(y_{2}^-,y_1^-) \right)
 \\
&+{1\ov i}\log\frac{ i^{Q}\,\Gamma\big[Q'-{i\ov
2}g\big(y_1^++\frac{1}{y_1^+}-y_2^+-\frac{1}{y_2^+}\big)\big]} {
i^{Q'}\Gamma\big[Q+{i\ov
2}g\big(y_1^++\frac{1}{y_1^+}-y_2^+-\frac{1}{y_2^+}\big)\big]}{1-
{1\ov y_1^+y_2^-}\ov 1-{1\ov
y_1^-y_2^+}}\sqrt{\frac{y_1^+y_2^-}{y_1^-y_2^+}} \,.~~~~~
\end{aligned}\eea
This formula represents the (logarithm of) improved bound state
dressing factor $\Sigma^{QQ'}$ of the mirror theory. It is
manifestly antisymmetric under interchanging the particles
$1\leftrightarrow 2$. Since for physical particles of the mirror
theory conjugation acts as $(y^{\pm})^*=1/y^{\mp}$, one can also
see that the dressing factor appears to be unitary. Indeed, the
last term in eq.(\ref{sigtot3}) is real, while to prove the
reality of the remaining terms one has to use the identities
(\ref{relP}) and (\ref{relPs}). Most remarkably, the factor
(\ref{sigtot3}) depends on the kinematic variables $y_i^\pm$ only,
which points to the uniqueness of the mirror S-matrix and,
therefore, to the validity of the whole mirror approach. This
factor also provides the final missing piece in derivation of the
TBA equations for the $\AdS$ mirror model \cite{AFmtba}.

\smallskip

Concerning the analytic properties of $\Sigma^{QQ'}$, the cuts
$\check{\rm x}^{(n)}_+$ in the region ${\cal R}_1$ are
cancelled\footnote{The same is true for the cuts $\check{\rm x
}^{(n)}_-$ in the region ${\cal R}_1$, as one could expect from
the validity of the crossing equations.} in the sum
$\Psi(y_{2}^+,y_1^+)+\Psi(y_{2}^-,y_1^+)$. Analogous cancellation
takes place in the other sums in eq.(\ref{sigtot3}). As a result,
the improved dressing factor $\Sigma^{QQ'}$ is a holomorphic
function in the intersection of the regions ${\cal R}_{1,1}$ and
${\rm Im}\, y_i^{\pm}<0$. We also note that $\Sigma^{QQ'}$ being
continued in the particle region will have there the cuts which
are analogues to those of the bound state dressing factor of
string theory  in the region ${\cal R}_{1,1}$.

\smallskip Finally, we mention that we have also made a
computation of the full mirror theory factor $\Sigma^{QQ'}$ by
using the bound state solution (\ref{sol1}) and verified that it
leads to the same result as the solution (\ref{sol2}).

\section{Conclusions}
In this paper we analyzed the analytic continuation of the  BES
dressing factor for a string bound state S-matrix on the
$z$-torus, and found its branch where the crossing equations are
satisfied. This provides a new test of the BES proposal.

\smallskip

In the particle region   the dressing factor of any bound state
S-matrix can be written in a universal form  in terms of a single
function $\chi$ of two variables.  Our results however show that
the analytic continuation and the choice of the branch of the
dressing factor depend on the type of scattered bound states, and
do not follow  just  from an analytic continuation of the function
$\chi$.

\smallskip

The analytic properties of the dressing factor  for fundamental
particles appear to be better than one could anticipate. In
particular, it is an analytic function in (the square  of) the
union of the particle region and the region Im$(x^\pm)<0$, and
regions obtained from this union by shifting it by $\om_2$ upward
or downward. The region Im$(x^\pm)<0$ was considered in
\cite{AFtba} as a candidate for the physical region of the mirror
model, and since the dressing factor is analytic in this region,
there might exist an integral representation for the factor,
similar to the DHM one, which makes analyticity manifest.

\smallskip

The string theory bound state dressing factor, however, is not
analytic in the region \mbox{Im$(x^\pm)<0$,}   and has  there a
finite number of cuts equal to $Q-1$ for a $Q$-particle bound
state. If $Q$ is even then one of the cuts is located on the real
momentum line of the mirror theory.
% The S-matrix of the mirror
%theory, on the other hand, should be unambiguously defined for
%real momenta.
We stress, however, that this discussion concerns the dressing
factor for bound states of string theory, but not the one for
bound states of the associated mirror model.

\smallskip

We have also determined the bound state dressing factor of the
mirror model. It is obtained by fusing the dressing factors of
mirror constituent particles that are located in various kinematic
regions of the $z$-torus and, for this reason, do not admit a
universal representation for their dressing factor in terms of
$\chi$-functions. Breakdown of this universality for particles
outside the particle region implies that the resulting bound state
dressing factor may depend on the choice of a bound state
solution. A $Q$-particle bound state equation has $2^{Q-1}$
different solutions, any of them can be used for constructing the
corresponding dressing factor leading, therefore, to a priory
different results. As we have shown, however, the factor
$\Sigma^{QQ'}$, which enters the TBA equations, does not suffer
from this ambiguity. The choice of a bound state solution is just
a matter of convenience and all $2^{Q-1}$ solutions lead to one
and the same physical S-matrix of the mirror theory.

\smallskip

Finally, we mention that the locations of all the cuts of  the
dressing factor depend only on a single variable $z_1$ or $z_2$
preserving the direct product structure of its domain, and this
hints at the existence of a  uniformizing variable such that the
dressing factor considered as a function of these two variables
becomes meromorphic.

%%%%%%%%%%%%%%%%%%%%%%%%%%%%%%%%%%%%%%%
\section*{Acknowledgements}
We would like to thank Zoltan Bajnok, Romuald Janik and Tomasz
Lukowski for interesting discussions, and Juan Maldacena for
useful comments on the manuscript. The work of G.~A. was supported
in part by the RFBR grant 08-01-00281-a, by the grant
NSh-672.2006.1, by NWO grant 047017015 and by the INTAS contract
03-51-6346. The work of S.F. was supported in part by the Science
Foundation Ireland under Grant No. 07/RFP/PHYF104.

\section{Appendix}

\subsection{Some results on the dressing phase for fundamental particles}
\label{sub:app1}

In this appendix we collect the auxiliary results on the analytic
continuation of the dressing phase for fundamental particles to
some other regions on the product of two infinite strips
$-{\om_1\ov 2}\le {\rm Im}(z)\le {\om_1\ov 2}$. These results can
be used, in particular, to verify the crossing equation which
arises upon shifting $z_2$ by $-\omega_2$.

%\vskip 0.7cm \noindent
%---------- FIGURE TOP ------------
\begin{figure}
\begin{minipage}{\textwidth}
\begin{center}
\includegraphics[width=0.45\textwidth]{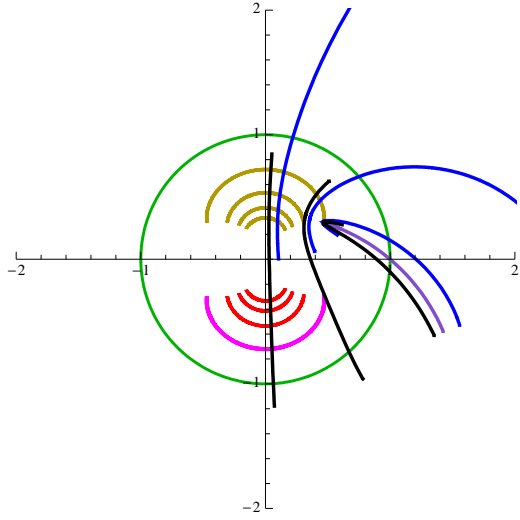}
\end{center}
\begin{center}
\parbox{5in}{\footnotesize{\caption{\label{funny_curves4} Blue and black curves represent the curves $x^-(z)$
going downward from the real line of the $z$ torus variable to the
line with Im$(z)=-{\om_2}/i$ and have $|$Re$(z)|\le {\om_1\ov 4}$,
and black curves $x^-(z)$ go downward and have $|$Re$(z)|\ge
{\om_1\ov 4}$. Any curve intersects the upper curve in the circle
and for the blue curves it is the first one they cross.
 }}}
\end{center}
\end{minipage}
\end{figure}
%---------- FIGURE END ------------
%\vskip -1.4cm

\subsubsection*{Region ${\cal R}_{0,-1}$:  $\{z_1,z_2\}\in {\cal R}_{0,-1}\ \ \Longrightarrow\ \   |x_1^\pm|>1\,;\  |x_2^+|>1\,,\  |x_2^-|<1$}

Let us now discuss very briefly the analytic continuation of the
dressing phase if the point $z_2$ is shifted downward, see Figure
11. The analytic continuation procedure we have already developed
basically repeats itself. Essentially, all we need to do is to
change $x_1^\pm\to x_2^\mp$, and to account properly for the
signs. Then, we get the following expressions for the functions
$\chi$ \bea\la{chimd1}
{\cal R}_{0,-1}:\quad \chi(x_1^\pm, x_2^-) &=& \Phi(x_1^\pm, x_2^-) + \Psi(x_2^-, x_1^\pm)\,,~~~~~~~~~~~~~~~~~~~\\
\nonumber \chi(x_1^\pm, x_2^+) &=& \Phi(x_1^\pm, x_2^+)\,. \eea

\subsubsection*{Region ${\cal R}_{0,-2}$:  $\{z_1,z_2\}\in {\cal R}_{0,-2}\ \ \Longrightarrow\ \   |x_1^\pm|>1\,;\  |x_2^+|<1\,,\  |x_2^-|<1$}
Shifting the point $z_2$ further downward into the anti-particle
region ${\cal R}_{0,-2}$, one gets \bea\la{chimd2b}
{\cal R}_{0,-2}:\quad \chi(x_1^\pm, x_2^-) &=& \Phi(x_1^\pm, x_2^-) + \Psi(x_2^-, x_1^\pm) +{ 1\ov i}\log \frac{x_1^\pm-{1\ov x_2^+}}{x_1^\pm-x_2^+}\,,~~~~\\
\nonumber \chi(x_1^\pm, x_2^+) &=& \Phi(x_1^\pm, x_2^+) +
\Psi(x_2^+, x_1^\pm)\,. \eea These formulae are used to check the
crossing equation (\ref{cr1b}).

\subsubsection*{Region ${\cal R}_{-1,-1}$:  $\{z_1,z_2\}\in {\cal R}_{-1,-1}\ \ \Longrightarrow\ \
|x_1^+|>1\, |x_1^-|<1;\  |x_2^+|>1\,,\  |x_2^-|<1$} This region
can also be considered as the one containing the real momentum line of the mirror model\footnote{Perhaps, one can use the terminology
``upper mirror" and ``down mirror" to distinguish the
regions obtained from the particle region by shifting the
corresponding $z$-variables upward or downward the real axis,
respectively.}, because it's symmetry axis is obtained from the
one of the particle region by shifting the latter downwards by the
quarter of imaginary period. To find the continuation of the
$\chi$-functions to this region, we can use the expressions for
$\chi$'s in ${\cal R}_{0,-1}$. We find
 \bea\nonumber
{\cal R}_{-1,-1}:\quad  \chi(x_1^+, x_2^+) &=& \Phi(x_1^+, x_2^+)
\\\nonumber
\chi(x_1^+, x_2^-) &=&  \Phi(x_1^+, x_2^-)+ \Psi(x_2^-, x_1^+)
\\\nonumber
\chi(x_1^-, x_2^+) &=& \Phi(x_1^-, x_2^+) - \Psi(x_1^-, x_2^+)\,,
\\
\chi(x_1^-, x_2^-)&=& \Phi(x_1^-, x_2^-)+ \Psi(x_2^-, x_1^-)-
\Psi(x_1^-, x_2^-)\\\nonumber &&~~~~~~~~~~~~~~~~~~~~+i
\log\frac{\Gamma\big[1+{i\ov
2}g\big(x_1^-+\frac{1}{x_1^-}-x_2^--\frac{1}{x_2^-}\big)\big]}
{\Gamma\big[1-{i\ov
2}g\big(x_1^-+\frac{1}{x_1^-}-x_2^--\frac{1}{x_2^-}\big)\big]}\,.\la{chirm1m1}
\eea
In fact, the dressing factor in this region can be easily obtained from the one in the region ${\cal R}_{1,1}$ by using the crossing equations with both arguments shifted by $-\om_2$. The dressing factors differ just by a simple factor of the form  ${x_1^+ x_2^-\ov x_1^- x_2^+}$.

\subsection{Identities for the $\Psi$-function}
\label{sub:app2} Here we present a list of identities satisfied by
the $\Psi$-function, which have been used in proving the crossing
equations for the dressing phase of both fundamental and mirror
particles.

\smallskip

For $|x_1^\pm|>1$ and $|x_2|>1$ the following identity is valid
\bea\la{id1} &&\Psi({1\ov x_1^-}, x_2) -\Psi({1\ov x_1^+}, x_2)=
\\ \nonumber &&~~~~~~~ ={1\ov i}\oint\frac{{\rm d }w}{2\pi i}
\log(w-x_2)\Big( w - {1\ov w}\Big)
%\\\nonumber
%&&~~~~~~~~~~~~~~~~~~~~~~~~~~~~~~~~~~~~~~~~~~\times
 \left[{1\ov (w-x_1^-)(w-{1\ov x_1^-})}+{1\ov (w-x_1^+)(w-{1\ov x_1^+})}\right]\\\nonumber
&&~~~~~~~=-{2\ov i}\log x_2 +{1\ov i}\log ({1\ov x_1^-}-x_2)({1\ov
x_1^+}-x_2)= {1\ov i}\log (1-{1\ov x_1^-x_2})(1-{1\ov x_1^+x_2})
\, .~~~ \eea

\noindent Analogously to the derivation above, we establish the
following identities

\begin{itemize}
\item[{\bf I}.] For $|x_1^+|<1\,,\  |x_1^-|>1 $ and $|x_2|>1$ one has
 \bea\la{id2}\quad \Psi({1\ov x_1^-}, x_2) -\Psi({1\ov
x_1^+}, x_2)&=& -{2\ov i}\log x_2 +{1\ov i}\log ({1\ov
x_1^-}-x_2)(x_1^+ -x_2)\, .~~~ \eea

\item[{\bf II}.] For  $|x_1^+|>1\,,\  |x_1^-|<1$ and $|x_2|>1$ one has
\bea\la{id3} \quad \Psi({1\ov x_1^-}, x_2) -\Psi({1\ov x_1^+},
x_2)&=&-{2\ov i}\log x_2 +{1\ov i}\log ( x_1^- - x_2)({1\ov x_1^+}
-x_2)\,
.~~~ \eea

\item[{\bf III}.] For $|x_1^+|>1\,,\  |x_1^-|>1$ and $|x_2|<1$ one has
\bea\nonumber \quad \Psi({1\ov x_1^-}, x_2) -\Psi({1\ov x_1^+},
x_2)&=& -\Psi({1\ov x_1^-}, {1\ov x_2}) +\Psi({1\ov x_1^+}, {1\ov
x_2}) +\Psi({1\ov x_1^-}, 0) -\Psi({1\ov x_1^+}, 0)~~~~~~~
\\\la{id4p}
&=& - {1\ov i} \log (x_1^--x_2)(x_1^+ -x_2) + i\log{g^2\ov 4}\,
.~~~ \eea

\item[{\bf VI}.] For $|x_1^+|<1\,,\  |x_1^-|>1$ and $|x_2|<1$ one has
\bea\nonumber && \Psi({1\ov x_1^-}, x_2) -\Psi({1\ov x_1^+}, x_2)=
-\Psi({1\ov x_1^-}, {1\ov x_2}) +\Psi({1\ov x_1^+}, {1\ov x_2})
+\Psi({1\ov x_1^-}, 0) -\Psi({1\ov x_1^+}, 0)
\\ \la{id4}
&&~~~~~=-\frac{2}{i}\log x_2 - {1\ov i} \log
(\frac{1}{x_1^-}-\frac{1}{x_2})(x_1^+ -\frac{1}{x_2})
+\frac{1}{i}\log\frac{x_1^+}{x^-_1}+ i\log{g^2\ov 4}\, .\eea In
deriving this formula we made use of the identity (\ref{id2})

\item[{\bf V}.] In proving the fourth relation we have  used the fact
that for $|x_1^+|>1$, \mbox{ $|x_1^-|>1$} and $|x_2|<1$ the
following identity holds
 \bea \la{id5} \quad \Psi({1\ov x_1^-}, 0) -\Psi({1\ov x_1^+},
0) &=&i\log (x_1^-x_1^+) + i\log{g^2\ov 4}\, ,~~~~~~~~~~~~~~ \eea
while in proving the fifth one for $|x_1^+|<1\,,\  |x_1^-|>1$ and
$|x_2|<1$ we relied on \bea \la{id5p} \quad \Psi({1\ov x_1^-}, 0)
-\Psi({1\ov x_1^+}, 0) &=&\frac{1}{i}\log \frac{x_1^+}{x_1^-} +
i\log{g^2\ov 4}\, .~~~~~~~~~~~~~~ \eea

\end{itemize}
\noindent Similar formulae exist for other values of $x_1^\pm$.

\subsection{An alternative derivation of the dressing factor}
\label{sub:app3} In this appendix we provide an alternative
derivation of the dressing factor for mirror bound states by
picking up a bound state solution with all constituent particles
being in the mirror region Im$(x_j^\pm)<0$, where
 \bea\la{mr1} {\rm
Im}(x_j^\pm) <0\,,\quad j=1,\ldots,Q\,,\quad |x_1^+| <1\,,\quad
|x_Q^-| >1\,. \eea Most compactly, the corresponding solution can
be written in terms of the function \bea\la{txu} \x(u) = {1\ov
2}\left(u - i\sqrt{4-u^2}\right)\,,\quad {\rm
Im}\big(\x(u)\big)<0\,, \eea as follows
 \bea\la{sol1}
x_j^-=\x\big(u+{i\ov g}(Q-2j)\big)\,,\quad x_j^+=\x\big(u+{i\ov
g}(Q-2j+2)\big)\,,\quad j=1,\ldots,Q\,.~~~~ \eea The function
$\x(u)$ satisfies the following inequalities \bea\la{txneq} |\x(u
- i y)| >1\,,\quad |\x(u + i y)| <1\,,\quad {\rm for}\ u\in
{\mathbf R} {\rm \ and}\ y>0\,.~~~~~ \eea Thus, we find for odd
$Q$ case, $Q=2m +1$ \bea &&|x_j^\pm|<1 \ {\rm\ if}\ j=1,\ldots, m\
\Rightarrow\ {\rm anti-particle\ region}\,,\\\nonumber
&&|x_{m+1}^-|>1\,,\ |x_{m+1}^+|<1\ \Rightarrow\ {\rm mirror\
region}\,,\\\nonumber &&|x_j^\pm|>1 \ {\rm\ if}\ j=m+2,\ldots, Q\
\Rightarrow\ {\rm particle\ region}\,. \eea So, only one particle
is in the mirror region.

\smallskip

If $Q$ is even, $Q=2m$, the story is more complicated and we find
\bea &&|x_j^\pm|<1 \ {\rm\ if}\ j=1,\ldots, m-1\  \Rightarrow\
{\rm anti-particle\ region}\,,\\\nonumber &&|x_j^\pm|>1 \ {\rm\
if}\ j=m+2,\ldots, Q\  \Rightarrow\ {\rm particle\
region}\,,\\\nonumber &&x_{m}^- = \x(u)\,,\ |x_{m}^+|<1\
\Rightarrow \left\{
\begin{array}{c}
{\rm anti-particle\ region\ if}\ x_{m}^- = \x(u+ i0)\cr {\rm
mirror\ region\ if}\ x_{m}^- = \x(u- i0)
\end{array}
\right. \,,\\\nonumber &&|x_{m+1}^-|>1\,,\ x_{m+1}^+ = \x(u) \
\Rightarrow \left\{
\begin{array}{c}
{\rm mirror\ region\ if}\ x_{m+1}^+ = \x(u+ i0)\, ,\cr {\rm
particle\ region\ if}\ x_{m+1}^+ = \x(u- i0)
\end{array}
\right. \,. \eea

 To apply the fusion procedure, as well as the formulae for
the analytic continuation, it is convenient to rewrite the
solution (\ref{sol1}) in terms of the function $x(u)$ introduced
in (\ref{xu}). This function satisfies the following inequalities
\bea\la{xneq} {\rm Im} \, x(u + i y) >0\,,\quad {\rm Im}\, x(u - i
y) <0\,,\quad {\rm for}\ u\in {\mathbf R} {\rm \ and}\  y>0\,.
\eea The relation between $\x(u)$ and $x(u)$ follows from
eqs.(\ref{txneq}) or (\ref{xneq}) \bea \la{xtx} \x(u - i y) = x(u
- i y)\,,\quad \x(u + i y) ={1\ov x(u + i y)}\,,\quad {\rm for}\
u\in {\mathbf R} {\rm \ and}\ y>0\,.~~~~~~ \eea Now we have two
separate cases for $Q$ odd and even.

\begin{itemize}

\item Thus, we find for odd $Q$ case, $Q=2m +1$ \bea\la{solQo}
\begin{aligned}
&x_j^-={1\ov x\big(u+{i\ov g}(Q-2j)\big)}\,,\quad x_j^+={1\ov
x\big(u+{i\ov g}(Q-2j+2)\big)}\,, \quad j=1,\ldots,
m\,,~~~~~~~~~\\ &x_{m+1}^-= x\big(u-{i\ov g}\big)\,,\quad
x_{m+1}^+={1\ov x\big(u+{i\ov g}\big)}\,,\\ &x_j^-= x\big(u+{i\ov
g}(Q-2j)\big)\,,\quad x_j^+= x\big(u+{i\ov g}(Q-2j+2)\big)\,,
\quad j=m+2,\ldots, Q\,.~~~~~~~~~
\end{aligned}\eea Thus, only the middle particle is in the region
${\cal R}_{1}$.

\smallskip

\item

If $Q$ is even, $Q=2m$, the structure of the solution is more
complicated \bea\la{solQe}\begin{aligned} &x_j^-={1\ov
x\big(u+{i\ov g}(Q-2j)\big)}\,,\quad x_j^+={1\ov x\big(u+{i\ov
g}(Q-2j+2)\big)}\,, \quad j=1,\ldots, m-1\,,~~~~~~~~\\
&x_m^+={1\ov x\big(u+{2i\ov g}\big)}\,, ~~\qquad\qquad x_{m}^- =
\left\{
\begin{array}{c}
{1\ov x(u+ i0)}\,,\ \ \ \   z_m\in {\rm anti-particle\ region}\cr
x(u- i0)\,,\  z_m\in{\rm mirror\ region}~~~~~~~~~~~~
\end{array}
\right. \,,\\ &x_{m+1}^-= x\big(u-{2i\ov g}\big)\,,\quad\qquad
x_{m+1}^+ = \left\{
\begin{array}{c}
{1\ov x(u+ i0)}\,,\ \ \ z_m\in {\rm mirror\ region}~\cr x(u-
i0)\,,\  z_m\in{\rm particle\ region}
\end{array}
\right. \,,\\ &x_j^-= x\big(u+{i\ov g}(Q-2j)\big)\,,\quad x_j^+=
x\big(u+{i\ov g}(Q-2j+2)\big)\,, \quad  j=m+2,\ldots, Q \,.
\end{aligned}\eea Again, only one particle is in the region
${\cal R}_{1}$.

\end{itemize}
For simplicity, we consider the case of $Q=2m+1$ and take the
second particle in the dressing factor to be a fundamental one of
string theory. Then, $m$ particles from the bound state solution
occur in the anti-particle region ${\cal R}_{2}$, $m$ particles
are in the particle region and one particle is in ${\cal R}_{1}$.
By using our analytic continuation formulae for dressing phases of
particles in the corresponding regions, we get \bea\la{thQ1}
\theta(y_1,y_2) &=&
 \Phi(y_1^+,y_2^+) - \Psi(y_1^+,y_2^+) - \Phi(y_1^+,y_2^-)+ \Psi(y_1^+,y_2^-)\\\nonumber
&-&\Phi(y_1^-,y_2^+) + \Phi(y_1^-,y_2^-)+{1\ov i}\log\prod_{j=1}^m
{ {1\ov x_j^-}-y_2^+\ov x_j^--y_2^+}{ x_j^--y_2^-\ov {1\ov
x_j^-}-y_2^-}\,. \eea Taking into account the formula
(\ref{Dthtot}), for the factor $\Sigma^{Q1}$ we find
\bea\la{thQ1bb} {1\ov i}\log\Sigma^{Q1}(y_1,y_2) &=&
 \Phi(y_1^+,y_2^+) - \Psi(y_1^+,y_2^+) - \Phi(y_1^+,y_2^-)+ \Psi(y_1^+,y_2^-)\\\nonumber
&-&\Phi(y_1^-,y_2^+) + \Phi(y_1^-,y_2^-)\\\nonumber &+& {1\ov
i}\log{ 1- {1\ov y_1^+y_2^-}\ov 1-{1\ov y_1^-y_2^+}}\prod_{j=1}^m
{ y_2^+\ov y_2^- }{ x_j^--y_2^-\ov x_j^--y_2^+ }\prod_{j=m+1}^{2m}
{ 1- {1\ov x_j^-y_2^-}\ov 1-{1\ov x_j^-y_2^+}}\,. \eea By using
eqs.(\ref{solQo}), this formula can be written in the following
form \bea\la{thQ1cc2} {1\ov i}\log\Sigma^{Q1}(y_1,y_2) &=&
 \Phi(y_1^+,y_2^+) - \Psi(y_1^+,y_2^+) - \Phi(y_1^+,y_2^-)+ \Psi(y_1^+,y_2^-)\\\nonumber
&-&\Phi(y_1^-,y_2^+) + \Phi(y_1^-,y_2^-)+ {1\ov i}\log{ 1- {1\ov
y_1^+y_2^-}\ov 1-{1\ov y_1^-y_2^+}}\prod_{j=1}^{Q-1} { 1- {1\ov
x\big(u+{i\ov g}(Q-2j)\big)y_2^-}\ov 1-{1\ov x\big(u+{i\ov
g}(Q-2j)\big)y_2^+}}\,, \eea which is obviously the same
expression as (\ref{thQ1c}). Considerations of the case $Q=2m$ is
analogous and leads to the same conclusion -- two different bound
state solutions produce one and the same dressing factor.

\subsection{Details on the derivation of the mirror bound state dressing factor}
\label{sub:app4}

In section 6 we constructed the dressing factor for bound states
of the mirror theory. This construction requires the use of
further identities for the $\Psi$-function which we present here.
Recall that the parameters $y_1^{+}=x_1^{+}$, $y_1^{+}=x_Q^{-}$
and $y_2^{+}=x_2^{+}$, $y_2^{+}=x_Q'^{-}$ of the mirror bound
states (\ref{bse}) satisfy the relations \bea \nonumber
y_1^{+}+\frac{1}{y_1^{+}}-y_1^{-}-\frac{1}{y_1^{-}}&=&\frac{2i}{g}Q\,
, \qquad
 y_2^{+}+\frac{1}{y_2^{+}}-y_2^{-}-\frac{1}{y_2^{-}
}=\frac{2i}{g}Q' \, . \eea Consider now a particular solution
(\ref{sol2}) of the bound state equations (\ref{bse}). If both
bound states involved in the construction of the dressing factor
are of this type, then \bea\la{ineqapp} |y_{1}^+|<1\, , \quad
|y_{1}^-|>1\, , \quad |y_{2}^+|<1\, \quad |y_{2}^-|>1\, . \eea By
using eq.(\ref{id1b}),  for this kinematic configuration we find
\bea\la{id1b2app} \hspace{-0.5cm}\Psi(y_{1}^+, y_2^-)
-\Psi(y_{1}^-, y_2^-)= {2Q\ov i}\log y_2^- -{1\ov
i}\log\prod_{j=1}^Q (w_j^+(y_1^-)-y_2^-)(w_j^-(y_1^+)-y_2^-)\,.~~~
\eea
 Taking into
account that for solution (\ref{sol2}) equations (\ref{gswj})
acquire the form
 \bea\begin{aligned} w_Q^+(y_1^-)& =
y_1^+=x_1^+=\frac{1}{x\big(u+{i\ov g}Q\big)}\,,
~~~~~~~w_j^-(y_1^+)={1\ov x_j^-}\, , \\
 w_Q^-(y_1^+) &={1\ov
y_1^-}=\frac{1}{x^-_Q}=\frac{1}{x\big(u-{i\ov g}Q\big)}\,,
 \end{aligned}\eea where $x^-_j$ is given by
eq.(\ref{sol2}), we obtain the following identity
\bea\la{id1b3app} \hspace{-0.8cm} \Psi(y_{1}^+, y_2^-)
-\Psi(y_{1}^-, y_2^-)= -{1\ov i}\log\left({y_1^+\ov y_2^-}-1
\right)\left({1\ov y_1^- y_2^-}-1
\right)\prod_{j=1}^{Q-1}\left(1-{1\ov x_j^-y_2^-}\right)^2. \eea
The latter formula also implies the following relation
\bea\la{id1b4app} \Psi(y_{2}^+, y_1^-) -\Psi(y_{2}^-, y_1^-)=
-{1\ov i}\log\left({y_2^+\ov y_1^-}-1 \right)\left({1\ov y_1^-
y_2^-}-1 \right)\prod_{k=1}^{Q'-1}\left(1-{1\ov
y_1^-z_k^-}\right)^2\,,~~~~~ \eea where $ z_k^- = 1/w_k^-(y_2^+) =
x(u_2 + {i\ov g}(Q'-2k))$. Further, by applying the basic identity
(\ref{relPs}) to eqs.(\ref{id1b3app}) and (\ref{id1b4app}), we
find the other two relations
 \bea\la{id1b5app} &&\Psi(y_{1}^+,
y_2^+) -\Psi(y_{1}^-, y_2^+)= {Q\ov i}\log {g^2\ov 4} -{1\ov
i}\log\prod_{j=1}^{Q-1}\left(1-{1\ov
x_j^-y_2^+}\right)^2\\\nonumber &&~~~~~~~~+{1\ov i}\log\left(y_2^+
-y_1^- \right)\left(y_2^+-\frac{1}{y_1^+}
\right)\prod_{j=1}^{Q-1}\left(u_1-u_2 +{i\ov
g}(Q-Q'-2j)\right)^2\,.~~~
\\\la{id1b6app} &&\Psi(y_{2}^+, y_1^+)
-\Psi(y_{2}^-, y_1^+)= {Q'\ov i}\log {g^2\ov 4} -{1\ov
i}\log\prod_{k=1}^{Q'-1}\left(1-{1\ov y_1^+
z_k^-}\right)^2\\\nonumber &&~~~~~~~~+{1\ov i}\log\left(y_1^+
-y_2^- \right)\left(y_1^+-\frac{1}{y_2^+}
\right)\prod_{k=1}^{Q'-1}\left(u_2-u_1 +{i\ov
g}(Q'-Q-2k)\right)^2\, ~~~ \eea valid for kinematic parameters
$y_{1,2}^{\pm}$ obeying inequalities (\ref{ineqapp}). In deriving
these formulae we have used the fact that \bea \nonumber\Psi({1\ov
y_1^+}, 0) -\Psi({1\ov y_1^-}, 0)&=&
\frac{Q}{i}\log\frac{g^2}{2}-\frac{1}{i}\log\Big[
w^+_Q(y_1^-)w^-_Q(y_1^+)\prod_{j=1}^{Q-1}(w_j^-(y_1^+))\Big]
\\
&=&
\frac{Q}{i}\log\frac{g^2}{2}-\frac{1}{i}\log\frac{y_1^+}{y_1^-}\prod_{j=1}^{Q-1}\frac{1}{(x_j^-)^2}\,
 \eea
and also the identity \bea (x_j^- -
y_2^-)\Big(1-\frac{1}{x_j^-y_2^-}\Big)=u_1-u_2+\frac{i}{g}(Q-Q'-2j)\,
, \eea where  $u_1$ and $u_2$ are the corresponding rapidity
parameters \bea\nonumber
u_1=x_1^{+}+\frac{1}{x_1^{+}}-Q\frac{i}{g}=x_1^{-}+\frac{1}{x_1^{-}}+Q\frac{i}{g}\,
,\quad
u_2=x_2^{+}+\frac{1}{x_2^{+}}-Q'\frac{i}{g}=x_2^{-}+\frac{1}{x_2^{-}}+Q'\frac{i}{g}\,
. \eea
 Finally, using the formulae
(\ref{id1b3app})-(\ref{id1b6app}), we find
 \bea\nonumber
-\Psi(y_1^+,y_2^+)&+&\Psi(y_1^+,y_2^-)+\Psi(y_{2}^+,y_1^+)-\Psi(y_{2}^+,y_1^-)
+{1\ov i}\log\prod_{j=1}^{Q-1} { 1- {1\ov x_j^-y_2^-}\ov 1-{1\ov
x_j^-y_2^+}} \prod_{k=1}^{Q'-1}{ 1- {1\ov y_1^+z_k^-}\ov 1-{1\ov
y_1^-z_k^-}} \\\nonumber &=& {1\ov
2}\left(-\Psi(y_1^+,y_2^+)+\Psi(y_1^+,y_2^-)+\Psi(y_{2}^+,y_1^+)-\Psi(y_{2}^+,y_1^-)
\right) \\ \la{prima!} &+&{1\ov
2}\left(-\Psi(y_1^-,y_2^+)+\Psi(y_1^-,y_2^-)+\Psi(y_{2}^-,y_1^+)-\Psi(y_{2}^-,y_1^-)
\right)
 \\\nonumber
&-&{1\ov i}\log{\prod_{j=1}^{Q-1}{g\ov 2}\left(u_1-u_2 +{i\ov
g}(Q-Q'-2j)\right)\ov \prod_{k=1}^{Q'-1}{g\ov 2}\left(u_2-u_1
+{i\ov
g}(Q'-Q-2k)\right)}+\frac{1}{2i}\log\frac{y_1^+y_2^-}{y_1^-y_2^+}\,.
\eea One more formula we need to complete the derivation is
 \bea
 \nonumber
&&{ \prod_{k=1}^{Q'-1}{g\ov 2}\left(u_2-u_1 +{i\ov
g}(Q'-Q-2k)\right)\ov \prod_{j=1}^{Q-1}{g\ov 2}\left(u_1-u_2
+{i\ov g}(Q-Q'-2j)\right)} \frac{\Gamma\big[1-{i\ov
2}g\big(y_1^++\frac{1}{y_1^+}-y_2^+-\frac{1}{y_2^+}\big)\big]}
{\Gamma\big[1+{i\ov
2}g\big(y_1^++\frac{1}{y_1^+}-y_2^+-\frac{1}{y_2^+}\big)\big]}\\
\la{gammaid}
&&~~~~~~~~~~~~~~~~~~~~~~~~~~~~~=i^{Q-Q'}\frac{\Gamma\big[Q'-{i\ov
2}g\big(y_1^++\frac{1}{y_1^+}-y_2^+-\frac{1}{y_2^+}\big)\big]}
{\Gamma\big[Q+{i\ov
2}g\big(y_1^++\frac{1}{y_1^+}-y_2^+-\frac{1}{y_2^+}\big)\big]}\, .
\eea

\end{document}